\documentclass[aps,prl,twocolumn,showpacs,superscriptaddress,reprint]{revtex4-2}
\usepackage{latexsym}
\usepackage{amssymb}
\usepackage{graphicx}
\usepackage{amsmath}
\usepackage{bm}
\usepackage[colorlinks,linkcolor=blue,citecolor=blue,urlcolor=blue]{hyperref}
\usepackage{verbatim}
\usepackage{mathrsfs}
\usepackage{extarrows}
\usepackage{comment}
\usepackage{mathtools,slashed}
\usepackage{times}

\usepackage{ulem}

\usepackage{cancel}
\usepackage{subfigure}
\usepackage{graphicx}
\usepackage{ragged2e}

\usepackage{placeins}

\usepackage{color}
\usepackage{siunitx}

\usepackage{ragged2e}

\newcommand*{\md}{\mathrm{d}}
\newcommand*{\mi}{\mathrm{i}}

\newcommand*{\ra}{\rangle}
\newcommand*{\tr}{\mathrm{Tr}}

\graphicspath{PRL_Reply_2025-07-11/}

\begin{document}

\title{Photomagnetic-Chiral Anisotropy mediated by Chirality-Driven Asymmetric Spin Splitting}

\author{Tianwei Ouyang}
\thanks{These authors contribute equally to this work.}
\affiliation{State Key Laboratory of Synergistic Chem-Bio Synthesis, School of Chemistry and Chemical Engineering, Frontiers Science Center for Transformative Molecules, Shanghai Key Laboratory for Molecular Engineering of Chiral Drugs, Shanghai Jiao Tong University, Shanghai 200240, China}
\author{Hang Su}
\thanks{These authors contribute equally to this work.}
\affiliation{School of Physics and Astronomy, Shanghai Jiao Tong University, Shanghai 200240, China}
\author{Wanning Zhang}
\affiliation{State Key Laboratory of Synergistic Chem-Bio Synthesis, School of Chemistry and Chemical Engineering, Frontiers Science Center for Transformative Molecules, Shanghai Key Laboratory for Molecular Engineering of Chiral Drugs, Shanghai Jiao Tong University, Shanghai 200240, China}
\author{Yingying Duan}
\affiliation{School of Chemical Science and Engineering, Tongji University, 1239 Siping Road, Shanghai, 200092, China}
\author{Yuxi Fang}
\email{sjtu15901600323@sjtu.edu.cn}
\affiliation{State Key Laboratory of Synergistic Chem-Bio Synthesis, School of Chemistry and Chemical Engineering, Frontiers Science Center for Transformative Molecules, Shanghai Key Laboratory for Molecular Engineering of Chiral Drugs, Shanghai Jiao Tong University, Shanghai 200240, China}
\author{Shunai Che}
\email{chesa@sjtu.edu.cn}
\affiliation{State Key Laboratory of Synergistic Chem-Bio Synthesis, School of Chemistry and Chemical Engineering, Frontiers Science Center for Transformative Molecules, Shanghai Key Laboratory for Molecular Engineering of Chiral Drugs, Shanghai Jiao Tong University, Shanghai 200240, China}
\author{Yizhou Liu}
\email{yizhouliu@tongji.edu.cn}
\affiliation{School of Physics Science and Engineering, Tongji University, 1239 Siping Road, Shanghai, 200092, China}

\begin{abstract}
Photomagnetic effects (PMEs), intrinsic to transition metals, arises from the interaction between light-induced angular momentum and electronic spin. These effects are suppressed in noble metals with high symmetry and electron density. Introducing chiral structures can induce photomagnetic-chiral anisotropy (PM-ChA) of metals by linking chirality and spin dynamics. However, a theoretical explain remains elusive. Here, we investigated the mechanism of PM-ChA in tetrahelix-stacked chiral nanostructured gold chains (CNACs) using first-principles calculations. Non-equilibrium Green’s function calculations reveal that chiral potentials enhance spin channel asymmetry by amplifying spin-orbit coupling (SOC)-induced spin splitting. Real-time time-dependent density functional theory simulations further identify SOC as the bridge connecting chiral spintronics to PME, where chirality-driven spin flips from asymmetric geometries generate opposing photomagnetic fields in materials of different handedness. These findings are consistent with experimental observations in chiral nanostructured gold films and provide a theoretical instruction for design metallic spintronic devices.
\end{abstract}

\date{\today}

%%%%%%%%%%%%%%%%%%%%

\maketitle

\textit{Introduction.} Photomagnetic effects (PMEs), where light influences magnetic properties, represent a captivating intersection of optics and magnetism \cite{enz1969photomagnetic, ohkoshi2001photo, lambert2014all}. While PMEs commonly exist in transition metal atoms through some spin-flipping processes, it is usually believed that noble metals with high electron density cannot be used as photomagnetic materials due to the absence of available spin polarized states \cite{wang2011molecular, foxley2023magneto}. From both fundamental science and practical application perspectives, it is highly desirable to unveil PMEs in noble metals through unconventional physical mechanisms. Chirality-induced spin selectivity (CISS) has emerged as a fundamental mechanism governing anisotropic charge and spin transport in chiral materials \cite{naaman2012chiral,bloom2024chiral,yang2021chiral}. CISS effect describes the asymmetric, spin-dependent electron transmission in chiral structures, leading to diverse chiral anisotropies \cite{duan2023chiral, ji2023spin}, including photo-induced~\cite{ding2021chiral}, electro-induced \cite{bai2021resistance}, and magnetic-induced chiral anisotropies \cite{zhang2024chirality}. Moreover, the application of chiral anisotropy has been extensively explored in various fields, including photocatalysis \cite{cui2023enantioselective}, electrocatalysis \cite{zhang2024chiral} and enantiomeric discrimination \cite{liu2020enantiomeric}. Notably, we found that photo-magnetic chiral anisotropy (PM-ChA) was particularly pronounced in noble metal systems \cite{liu2022photomagnetic, yang2022chiral}. 

In chiral gold nanostructures, PM-ChA emerges through laser-driven electron dynamics along chiral trajectories. Relativistic spin-orbit coupling (SOC) converts structural chirality into orbital angular momentum (OAM), generating handedness-dependent transient magnetic dipoles \cite{ray1999asymmetric, shitade2020geometric, guo2012spin, zhao2025magnetochiral}. The chiral potential induces momentum-selective spin polarization through OAM-spin locking \cite{sahu2021effect, evers2022theory, yang2023monopole}, creating asymmetric spin population amplified by optical Stark splitting and geometric phase effects \cite{zeng2025photo}. Spin-polarized currents self-induce local magnetic fields that feed back into the system via chiral Lorentz forces and spin-dependent scattering, dynamically reinforcing the photomagnetic response~\cite{vinas2022microscopic,matsubara2022polarization,dainone2024controlling}. This synergy between structural chirality and SOC breaks noble metals' inherent symmetry, enabling more than 50\% spin polarization within 200 fs - faster than electron-lattice equilibration \cite{siegrist2019light, adhikari2023interplay, yu2020chirality}. The cascading effects of orbital polarization, spin-momentum locking, and self-amplifying magnetic feedback establish light-controlled spin population through chiral relativistic interactions.

To elucidate the microscopic origin of PM-ChA, we establish atomic-scale models of chiral gold nanostructured chains (CNACs) as experimentally observed one-dimensional helical assemblies of tetrahedral units. Through multi-scale quantum simulations integrating static electronic structure and real-time dynamic propagation, we resolve SOC mediated light-matter interactions that drive spin-polarized transport. This approach allows us to capture the SOC-induced coupling between optical fields and electronic states, revealing spin-selective transport behavior consistent with experimental observations~\cite{liu2022photomagnetic}. These findings advance the manipulation of spin-photon correlations in chiral nanoscale systems, offering new design principles for chiral opto-spintronic devices.

\begin{figure*}[tbp]
\centering
\includegraphics[width=0.735\textwidth]{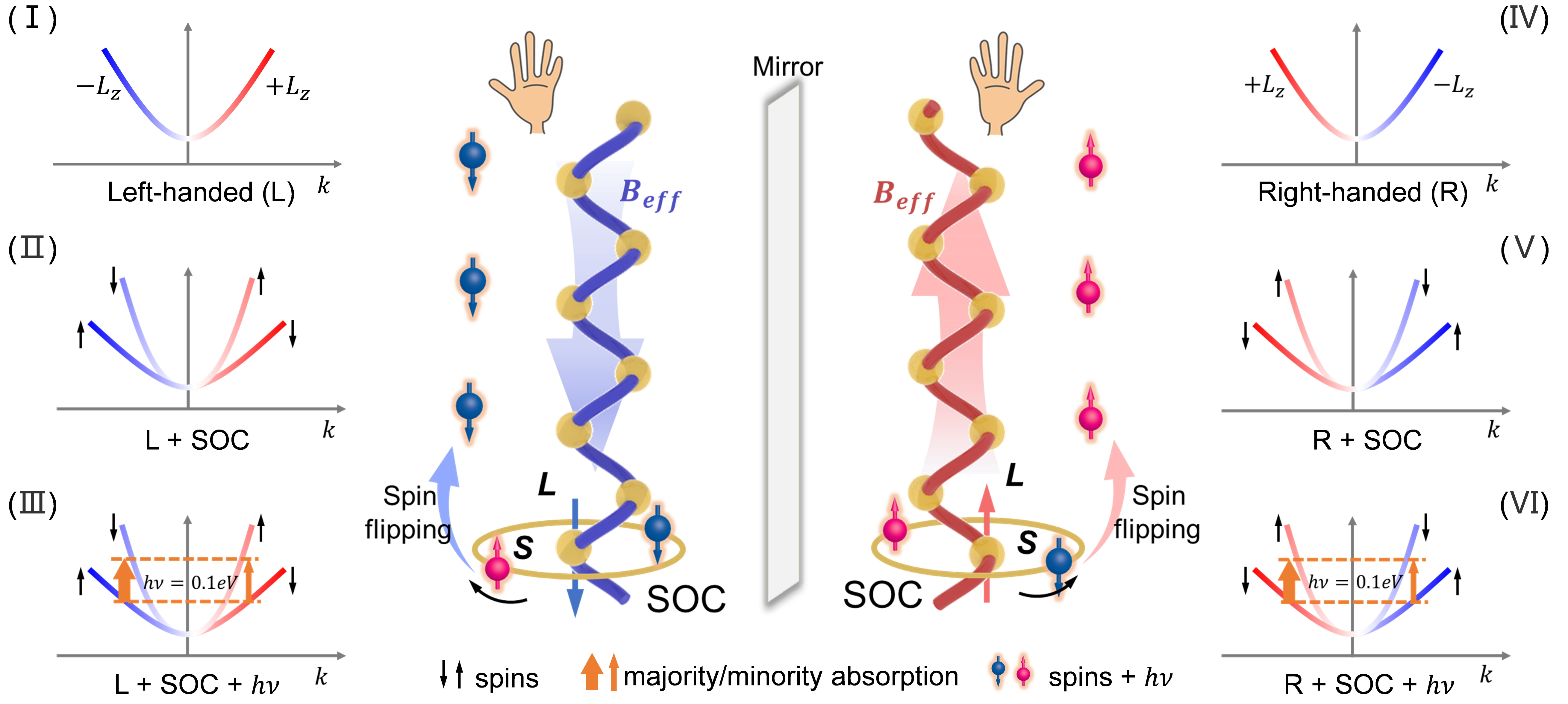}
\caption{\label{fig:1} Mechanism of PM-ChA. Schematic of spin dynamics in chiral materials under laser irradiation. Left-handed (in blue) and right-handed (in red) helical configurations generate chirality-dependent band splitting through geometric symmetry breaking. (I) Ground-state orbital texture of left-handed system shows antiparallel orbital polarization $L_z$ distribution (red/blue bands) induced by chiral symmetry breaking. (II) SOC-induced band splitting in left-handed system maintains parallel $L_z$ alignment in $\pm\bm{k}$ branches, creating spin population asymmetry ($\uparrow$: spin up, $\downarrow$: spin down). (III) Under laser excitation (orange arrow: $h\nu$ absorption) the absorption channel at $\pm \bm{k}$ with direct band gap $h\nu$ is allowed due to energy conservation. Because of the conservation of momentum, the channel close to the wave vector of incident light provides the majority absorption channel which is dependent on the incident {\color[rgb]{1,0.5,0}\sout{light direction}} wave vector \cite{wan2023anomalous}. (IV-VI) right-handed system exhibits mirrored behavior with reversed $L_z$ alignment asymmetry and $+\bm{k}$-directed spin current.
}
\end{figure*}

\textit{Mechanism of PM-ChA.} PM-ChA originates from the synergistic interplay of chirality-induced orbital momentum locking, SOC, and asymmetric photoexcitation (Fig. \ref{fig:1}). The chiral geometry breaks the inversion symmetry, polarizing electronic bands with opposite OAM textures ($L_z\neq 0)$ in left- and right-handed systems [Figs.~\ref{fig:1}(I, IV)]. SOC interaction $H_{SOC} \propto \bm{L} \cdot \bm{S}$ lifts spin degeneracy through OAM-dependent splitting, generating asymmetric spin distributions [Figs.~\ref{fig:1}(II, V)]. Optical excitation drives dual photomagnetic responses through distinct pathways~\cite{qiu2023axion, matsubara2022polarization}. Photon absorption ($h\nu$) initiates time-reversal symmetry breaking, generating nonequilibrium electronic states, while structural chirality governs asymmetric interband transitions at $\pm \bm{k}$ points that produce net spin polarization [Figs.~\ref{fig:1}(III, VI)]. The channel closer to the wave vector of incident light is the majority absorption channel which depends on the incident direction of light. The direction-dependent optical phenomenon ($e.g.$ direction-constraint circular dichroism) has been discovered in chiral materials recently~\cite{wan2023anomalous}. The broken inversion symmetry inherent to chiral geometries dictates transition probability asymmetries, with SOC mediating spin-flip processes that amplify these chiral PMEs. Crucially, left- and right-handed configurations exhibit reversed spin current polarities, a direct consequence of their geometrically mirrored OAM textures and structural handedness. While phonon-induced spin polarization dominates in insulating chiral systems (e.g., $\alpha$-quartz)~\cite{kim2023chiral, funato2024chirality, ohe2024chirality}, electron-electron correlations prevail in metallic CNACs due to their high carrier density and large atomic mass, rendering electron-phonon couplings negligible.

The PM-ChA mechanism fundamentally relies on three coupled factors: 
chiral symmetry breaking enables OAM polarization, SOC converts orbital momentum to spin polarization, and optical fields drive directional spin currents through selective photon angular momentum transfer. 
This trinity of geometric chirality, relativistic interactions, and light-matter coupling establishes handedness-dependent photomagnetic responses in chiral systems. 

\begin{figure*}[tbp]
\centering
\includegraphics[width=0.78\textwidth]{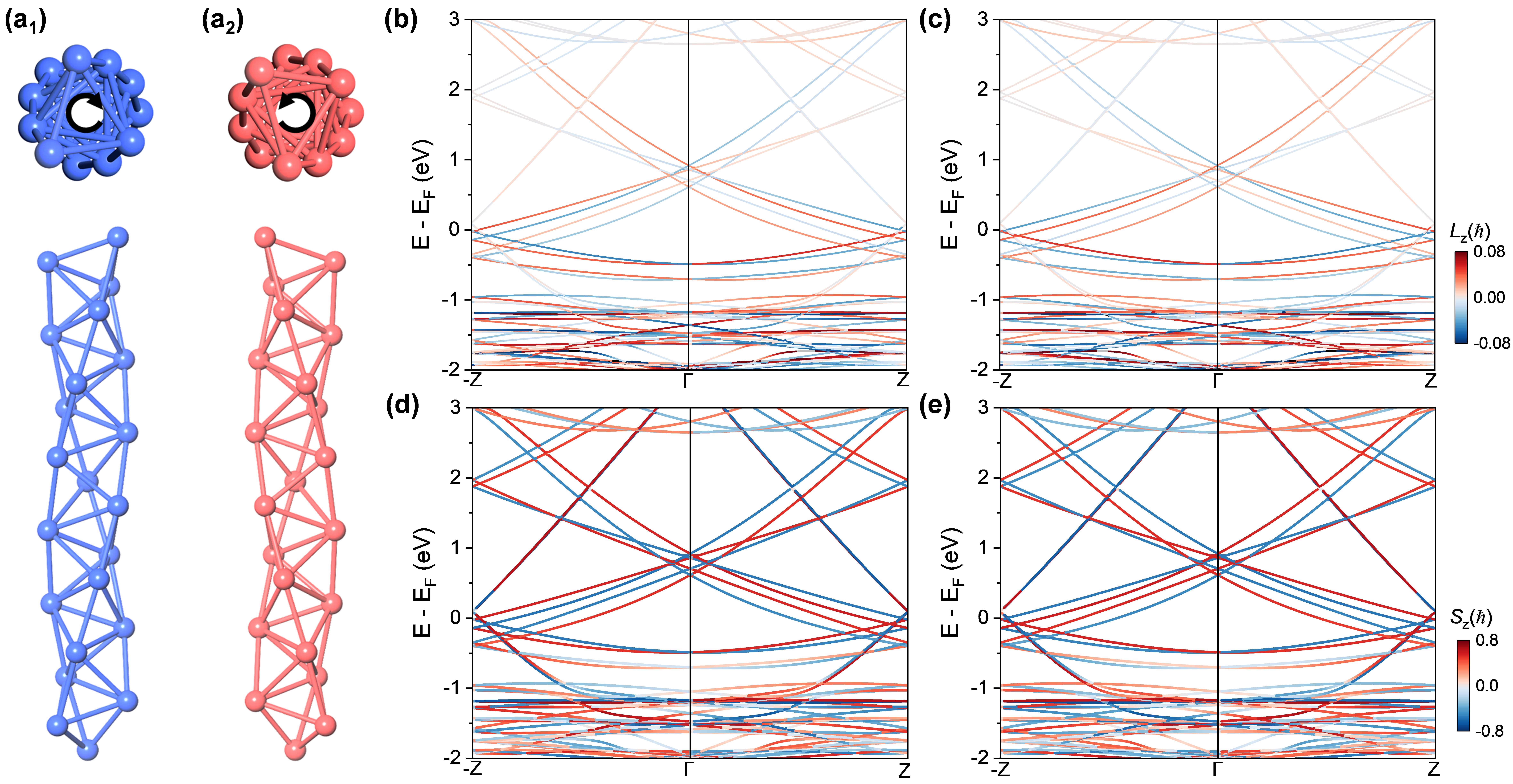}
\caption{\label{fig:2} Orbital texture and spin population. (a) Top and side views of the atomic structures of L-CNAC ($\text{a}_1$) and R-CNAC ($\text{a}_2$). Chirality arises from the clockwise (left-handed) or counterclockwise (right-handed) stacking orientation of tetrahedra. (b, c) The \textit{ab-initio} band structure of L-CNAC (b) and R-CNAC (c) with orbital texture. The orbital texture refers to the parallel or antiparallel relation between orbital polarization $L_z$ and the momentum. (d, e) Spin-resolved band structures of L-CNAC (d) and R-CNAC (e), respectively. Spin projection ($S_z$) is color-coded: red (blue) indicates spin-up (spin-down).
}
\end{figure*}

\textit{Chiral geometry and asymmetric spin population.} Chiral nanostructured Au chains (CNACs), formed by helical assemblies of tetrahedra (Boerdijk-Coxeter-Bernal helices), exhibit intrinsic chirality and broken translational and rotational symmetry \cite{zhu2014chiral, liu2020enantiomeric}. These quasi-1D systems exist as left-handed (L-CNAC) and right-handed (R-CNAC) enantiomers [Fig.~\ref{fig:2}(a)]. Their incommensurate periodicity complicates first-principles modeling, prompting our investigation of a commensurate analog with twisted atomic configurations along the $z$-axis [see Supplementary Materials~\cite{Append} 
% \textcolor[rgb]{1,0.5,0}{"Twisting modification of Chiral Nanostructured Au Chains" section}
Eqs.(S3) and (S4)]. The inherent lack of inversion symmetry in CNACs enables strong SOC effects that critically influence their electronic properties.

SOC induces substantial modifications in CNACs' electronic structures through spin-orbital interaction. Band structure comparisons [Fig. S2(a)] reveal that SOC splits degeneracies near Fermi level $(E_F)$ at $\Gamma$ and $\pm Z$ points, particularly through spin-momentum coupling that alters band crossing behaviors. This splitting correlates with enhanced density of states (DOS) peaks near $E_F$ [Fig. S2(b)], suggesting SOC-mediated state redistribution crucial for chiral optical responses. Projected DOS analysis (Fig. S3) shows minimal SOC impact on $s$-orbitals but significant modifications to $p$-orbitals and $d$-orbital enhancement near $E_F$. The strong $d$-orbital response originates from their high intrinsic angular momentum \cite{gao2024giant, kim2024orbital}, while structural chirality facilitates asymmetric SOC coupling through broken inversion symmetry.

A key feature emerges in the $L_z$, representing atomic OAM. The non-quantized $L_z$ exhibits opposite signs at $+Z$ and $-Z$, i.e., $L_z (\bm{k})=-L_z (-\bm{k})$ [Figs. \ref{fig:2}(b, c)], maintaining time-reversal symmetry while generating orbital polarization analogous to Weyl fermion chirality~\cite{wang2023quantum}. This counter-propagating orbital texture enables orbital polarization effects~\cite{liu2021chirality} through momentum-dependent OAM reversal.

Notably, SOC induces chirality-dependent spin splitting despite zero net magnetization. Spin-resolved band structures demonstrate enantiomeric inversion of $S_z$ projections between L-CNAC and R-CNAC [Figs. \ref{fig:2}(d, e)], establishing collinear spin-momentum locking along the chain axis \cite{zhao2023chirality}. Such spin population asymmetry is absent in achiral Au chains (ANACs, Fig. S4), confirming chirality's essential role in spin-state modification. The large SOC strength of Au atoms produces $S_z$ magnitudes significantly exceeding $L_z$ values, with spin polarization being geometry-sensitive through lattice parameter dependence. These characteristics position CNACs as promising platforms for investigating chirality-induced spin selectivity and low-dimensional spin transport phenomena.

\begin{figure*}[tbp]
\centering
\includegraphics[width=0.98\textwidth]{FIG3-1.jpg}
\caption{\label{fig:3} Time evolution of chiral photomagnetic responsiveness with SOC. (a-c) Time-dependent magnetic moment components under a uniform external electric field for L-CNAC (a), R-CNAC (b), and ANAC (c), respectively. (d-f) Time-dependent magnetic moment components driven by linearly polarized light for L-CNAC (d), R-CNAC (e), and ANAC (f), respectively. (g-i) Time-dependent occupations of spin-up and spin-down channels driven by linearly polarized light for L-CNAC (g), R-CNAC (h), and ANAC (i), respectively. The $x$, $y$, and $z$ directions correspond to periodic boundary conditions for $z$-axis. Light excitation at 12400 nm (0.1 eV) with $1.0 \times 10^{12}$ W/cm$^2$ intensity (electric field magnitude $\sim2.74 \times 10^9$ V/m), 300 fs Gaussian pulse, and $z$-axis polarization drives the photomagnetic response. }
\end{figure*}

\textit{Spin selectivity of CNACs.} We have developed a quantum transport theory for chirality-induced spin polarization by integrating first-principles density functional theory with the non-equilibrium Green's function formalism (DFT-NEGF) \cite{PhysRevB.65.165401, PhysRevLett.68.2512, naskar2023chiral}. The electrode-central region-electrode configuration preserves strict continuity of chiral symmetry [Fig. S9], enabling precise Bloch wavefunction matching at the interface that eliminates quantum reflections induced by potential barriers. Our computational framework directly solves spin-resolved Kohn-Sham equations under non-equilibrium conditions, where SOC effects naturally arise from the chiral potential gradient operator in the Hamiltonian \cite{Append}. Calculated current-voltage characteristics show quantitative agreement with experimental magnetic-tip conducting atomic force microscopy (mc-AFM) measurements of spin selectivity \cite{liu2022photomagnetic}.

Figures S10(a) and (b) present the current-voltage characteristics of L-CNAC and R-CNAC, respectively. Notably, the spin-down channel in L-CNAC and the spin-up channel in R-CNAC demonstrate significantly higher conductance compared to their respective counterpart spin channels, revealing chirality-dependent spin polarization in charge transport. This spin-selective behavior is further corroborated by the temporal evolution of total magnetic moment under external electric field stimulation [Figs.~\ref{fig:3}(a) and (b)]. Both enantiomers exhibit mirror-symmetric magnetic moment dynamics, converging to an average value of $0.11 \times 10^{-5}$ $\mu_B$ at $5.6$ fs. The observed anti-phase oscillations in magnetic moments suggest that spin-polarized electron transport through opposite-handed CNACs generates mutually opposing magnetic fields.

Detailed analysis reveals that the negative magnetic moment in L-CNAC corresponds to an effective magnetic field preferentially facilitating spin-down electron transport, whereas R-CNAC's positive magnetic moment indicates preferential spin-up transport. In contrast, the achiral ANAC control system displays complete absence of magnetic moment components [Fig.~\ref{fig:3}(c)], confirming chirality as the essential factor governing spin selectivity. The underlying mechanism of chirality-induced spin polarization is illustrated in Figs. S10 (c$_1$) and (c$_2$). When electrons traverse CNACs from left to right electrodes, the molecular geometry-dependent effective magnetic field acts as a spin filter: L-CNAC promotes spin-down electron transmission while R-CNAC enhances spin-up transmission to the right electrode.

\textit{Time evolution of chiral photomagnetic responsiveness.} The origin of CISS effects in transport phenomena remains debated, with two principal mechanisms proposed: spin-state filtering and spin-flip transitions. To elucidate their relative contributions in chirality-locked antisymmetric transport, we investigate ultrafast spin dynamics mediated by asymmetric SOC in CNACs under optical excitation using real-time time-dependent DFT (rt-TDDFT) simulations \cite{theilhaber1992ab, yabana1996time, xu2024real, siegrist2019light}. Figures \ref{fig:3} (d)-(i) reveal that a linearly polarized 12400 nm laser pulse induces $z$-axis-aligned in-plane polarization dynamics within 300 fs. Through joint density of states analysis Fig. S11, this infrared wavelength was specifically selected to overcome the limited free-electron transition efficiency in metallic CNACs under visible light excitation.

Correlated magnetic moment dynamics in Figs. \ref{fig:3}(d) and (e) reveal chirality-dependent signatures. L-CNAC's $M_z$-evolution displays pronounced asymmetry compared to R-CNAC, culminating in stable unidirectional spin polarization post-160 fs. This establishes photomagnetic field generation through laser-induced spin alignment. Crucially, the enantiomeric pair exhibits identical response magnitudes with opposite polarization directions, confirming chirality as the symmetry-breaking origin.

Figures \ref{fig:3}(g) and (h) demonstrate time-resolved spin population differences ($\Delta n_{\alpha=\uparrow}$, $\Delta n_{\alpha=\downarrow}$) for L-CNAC and R-CNAC under intense laser illumination. Photon absorption initiates plasmon-driven transitions that persist until magnetic stabilization at approximately 160 fs. Mirror-image responses emerge: L-CNAC exhibits spin-up to spin-down conversion while R-CNAC displays the inverse process. These dynamics, quantified through $\langle S_z(t) \rangle$ computations~\cite{neufeld2023attosecond}, directly manifest symmetry-breaking effects in chiral structures under optical excitation. The observed spin-state asymmetry confirms chirality's governing role in SOC-mediated spin-flip processes.

These findings bridge spin-selective transport and photomagnetic phenomena. Asymmetric SOC modifies spin-flip probabilities between channels, enabling chirality-dependent photomagnetic responses \cite{ryan2023optically}. Comparative analysis with achiral ANAC [Figs. \ref{fig:3}(f) and (i)] confirms the absence of both spin selectivity and photo-magnetism under identical excitation, verifying chirality's essential role. Complementary evidence from SOC-absent simulations Fig. S15 shows $10^6$-fold photomagnetic suppression and identical L/R-CNAC responses within computational noise. This dual verification establishes SOC as the critical mediator between chiral structures and spin-photon coupling.

The amplified photomagnetic response in chiral systems stems from SOC-enabled cooperative effects: optical fields induce non-equilibrium electron distributions that drive synchronized spin-flip transitions across channels. This cooperative mechanism creates positive feedback between spin polarization and photomagnetic enhancement. Conversely, SOC removal decouples spin-photon interactions, eliminating both symmetry breaking and magnetic response modulation. Our results demonstrate that chiral geometry provides the structural asymmetry while SOC enables the quantum mechanical coupling essential for photo-spintronic functionality.

\textit{Conclusions.} Through first-principles quantum dynamical simulations, we elucidate the quantum mechanical mechanism underlying PM-ChA in tetrahelix-stacked CNACs. Our DFT-NEGF framework reveals that chiral potential gradients induce asymmetric SOC, creating spin-polarized transport channels through symmetry-broken spin population distributions. The rt-TDDFT simulations demonstrate that light-driven SOC mediates enantiomer-specific spin-flip dynamics, generating oppositely oriented photomagnetic fields in left- and right-handed CNACs. These results establish SOC as the fundamental quantum mediator linking chiral geometry to photomagnetic phenomena. The synergistic interplay between structural chirality and SOC enhances light-matter interaction cross sections by orders of magnitude, producing giant PM-ChA signals. Crucially, polarized light excitation creates non-equilibrium spin distributions that undergo chirality-locked flipping transitions, resolving the long-standing debate about the origin of photomagnetic handedness dependence. This theoretical breakthrough provides a design roadmap for chiral quantum materials with programmable spin-photon coupling, advancing development of chiral optoelectronics and spin-based photonic devices.

\begin{acknowledgments}
\textit{Acknowledgements---}This work was supported by the National Key R\&D Program of China (Grant No. 2021YFA1200301, S.C.), the National Natural Science Foundation of China (Grant No. 12404279, Y.L.), 
Shanghai Pilot Program for Basic Research-Shanghai Jiao Tong University (21TQ1400219), the Innovation Program for Quantum and Technology (Grant No. 2023ZD0300500, Y.L.) and Shanghai Pujiang Program (Grant No. 23PJ1413000, Y.L.).
\end{acknowledgments}

%\bibliographystyle{apsrev4-1}
%\bibliography{PRL_Reply_2025-07-11/refs}


\newpage
\onecolumngrid 
\appendix
\setcounter{figure}{0}
\setcounter{equation}{0}
\renewcommand\thefigure{S\arabic{figure}}
\renewcommand\theequation{S\arabic{equation}}

\section{\large{} Supplemental Materials on ``Photomagnetic-Chiral Anisotropy mediated by Chirality-Driven Asymmetric Spin Splitting''}

% \maketitle

\subsection{Twisting modification of Chiral Nanostructured Au Chains}

The Chiral Nanostructured Au Chains (CNAC) we examine is a purely mathematic model of a one-dimensional (1D) chiral spin chain, formed by linear stacking structure of regular tetrahedra \cite{zhu2014chiral, liu2020enantiomeric, liu2022photomagnetic}. This arrangement is known as the Boerdijk-Coxeter helix \cite{boerdijk1952some, coxeter1993regular}, or the Boerdijk-Coxeter-Bernal helix \cite{bernal1960geometry}. In this structure, each tetrahedron’s dihedral angle-the angle between two faces is given by $\arccos \left(1/3\right) \simeq{70.53}^\circ$. A tetrahedron in the CNAC rotates about the c-axis by an angle of $\alpha_0 = \arccos(2/3) \simeq 131.71^{\circ}$. 

Considering a helix with Cartesian coordinates for each vertex $V_i = (x_i, y_i, z_i)$ in three-dimensional space, then coordinate of the $N$-th vertex starting from this vertex $V_i$ is
\begin{equation}
  V_{i+N} = (x_{i+N}, y_{i+N}, z_{i+N}) = (l r \cos (N \alpha_0), \pm l r \sin (N \alpha_0), z_i + N l h_0),
\end{equation}
where $r$ and $l$ represent the radius of cylinder containing the vertices and the unit length of each tetrahedral, respectively. The signs $+$ and $–$ represent right- and left-handedness, respectively. The constants are given by 
\begin{equation}
  \alpha_0 = \arccos (-2/3) \approx 131.8^{\circ} \ \ \ , \ \ \ h_0 = \frac{1}{\sqrt{10}} \ \ \ , \ \ \ r = \frac{3 \sqrt{3}}{10}.
\end{equation}

Since the $\alpha_0$ is an incommensurate angle ($i.e.$, it does not equal any rational multiple of $180^\circ$), we introduce a small twisting angle $\theta$ to make the structure periodic and commensurate without loss of the general physical picture. After twisting, the coordinates are evaluated as
\begin{equation}
  V'_{i+N} = (x'_{i+N}, y'_{i+N}, z'_{i+N}) = (l r \cos (N \theta_0), \pm l r \sin (N \theta_0), z_i + N l h_0) = (x_i, y_i, z_i + N h_0),
\end{equation}
where the new angle $\theta_0 = \alpha_0 + \theta$ ensures the commensurate relationship
\begin{equation}
  N (\alpha_0 + \theta) = M \cdot (2 \pi) \ \ \ , \ \ \ N, M = 1, 2, 3, \cdots .
\end{equation}

Several positive integer solutions for $N$ and $M$ are enumerated as follows:
\begin{center}
\begin{tabular}{c|c|c}
  Sequence & $\theta$ & $(M, N)$ \\
  \hline
  \#1 & $-11.80^{\circ}$ & (1, 3) \\
  \#2 & $+3.19^{\circ}$ & (3, 8) \\ 
  \#3 & $-0.90^{\circ}$ & (4, 11) \\ 
  \#4 & $-3.24^{\circ}$ & (5, 14) \\ 
  \#5 & $+0.82^{\circ}$ & (7, 19) \\
  \#6 & $-2.21^{\circ}$ & (9, 25) \\
  \#7 & $+1.52^{\circ}$ & (10, 27) \\
  \#8 & $+0.19^{\circ}$ & (11, 30) \\
\end{tabular}
\end{center}
In choosing the twisting angle $\theta$, it is practical to select the smallest possible angle to minimize computational cost, while ensuring a manageable structure. Based on these considerations, we select solution \#3, applying a twist of ${-0.90}^\circ$ per tetrahedron, resulting in a periodic structure with 11 atoms per unit cell. 

\subsection{Band Structure calculation based on Wannier Hamiltonian}
\subsubsection{Band Structure calculation and Wannier Hamiltonian}

We use Vienna \textit{ab-initio} Simulation Package (VASP) to analyze energy band structure for our golden chain with CNAC configuration, and employ Wannier90 to construct tight-binding Hamiltonian form in the Wannier basis. In details, the so-called Wannier Hamiltonian can be written as
\begin{equation}
	H_{\text{Wannier}} = \sum_{i,\alpha,\sigma=\uparrow,\downarrow}\sum_{i^\prime,\alpha^\prime,\sigma^\prime=\uparrow,\downarrow}{{\ t}_{ii^\prime,\alpha\alpha^\prime,\sigma\sigma^\prime}c_{i,\alpha,\sigma}^\dag c_{i^\prime,\alpha^\prime,\sigma^\prime}}+t_{i^\prime i,\alpha^\prime\alpha,\sigma^\prime\sigma}c_{i^\prime,\alpha^\prime,\sigma^\prime}^\dag c_{i,\alpha,\sigma}
\end{equation}
where $i,\alpha,\sigma$ ($i^\prime,\alpha^\prime,\sigma^\prime$) represent the indices of site coordinate, orbits and spin, respectively. Here, $t_{ii^\prime,\alpha\alpha^\prime,\sigma\sigma^\prime}$ is the hopping coefficient from the site $i$, orbital $\alpha$ ($\alpha$ = $s$, $p$, $d$ for Au), and spin $\sigma$ to that for site $\alpha$, orbital $\alpha$, and spin $\sigma^\prime$, and $t_{i^\prime i,\alpha^\prime\alpha,\sigma^\prime\sigma}$ describes the reverse hopping. 

Typically, the Wannier Hamiltonian can be computed straightforwardly and accurately by using Wannier90. However, the complicated structure of the BCB helix in our study poses challenges, as it does not allow for a Wannier Hamiltonian that preserves certain symmetries, most importantly, time-reversal symmetry (TRS) because there are not any magnetizations in any golden atom arrays. To address this, we introduce a transformation called time-reversal symmetrization for the computed Wannier Hamiltonian.

\subsubsection{Time-reversal symmetrized operation for Wannier Hamiltonian}

The Time-reversal symmetry (TRS) operators is defined as~\cite{tasaki2020physics}
\begin{equation}
	\hat{\Theta} = \hat{\tau} \hat{K} \ \ \ , \ \ \ \hat{K} f(r) = f^{*} (r) \hat{K},
\end{equation}
Here, $\hat{K}$ is antiunitary operators, and we typically use the Pauli matrix $\sigma^y=\left(\begin{matrix}0&-\mi\\ \mi&0\\\end{matrix}\right)$ as part of $\hat{\tau}$. It is important to note that TRS obeys
\begin{equation}
	\hat{\Theta} H \hat{\Theta}^{-1} = H
\end{equation}
both in momentum space and in real space. In momentum space, the energy band is symmetric about the $\Gamma$-point, while in real space, additional considerations are necessary later. 

In real space case, one can seperate it into spinless orbital and spin patrs for Hamiltonian in Wannier basis. For the orbital functions, the time-reversal transformation is to change into its complex conjugate based on the properties of spherical harmonic function $Y_{l}^{m} (\theta, \phi)$, where $\theta$ and $\phi$ are polar and azimuthal angles in spherical coordinates, and $l$ and $m$ denote orbital and magnetic quantum numbers, respectively~\cite{sakurai2020modern}. Under TRS, the spherical harmonic transforms as follows: 
\begin{equation}
	Y_{l}^{m} (\theta, \phi) \rightarrow Y_{l}^{m *} (\theta, \phi) = (-1)^m Y_{-l}^{m} (\theta, \phi) .
\end{equation}
Thus, the states $|l, m\rangle$ transforms as
\begin{equation}
	\hat{\Theta} | l, m \ra = (-1)^m | l, -m \rangle. 
\end{equation}

Applying this symmetrized formula to the $s$-, $p$-, $d$-orbital wave functions (which are the focus in our CNAC study), 
\begin{subequations}
\begin{align}
	\psi_s &= Y_{0}^{0} \\
	\psi_{p_x} &= \frac{1}{\sqrt{2}} \left( Y_{1}^{-1} - Y_{1}^{1} \right), \\
	\psi_{p_y} &= \frac{\mi}{\sqrt{2}} \left( Y_{1}^{-1} + Y_{1}^{1} \right), \\ 
	\psi_{p_z} &= Y_{1}^{0}, \\
	\psi_{d_{xz}} &= \frac{1}{\sqrt{2}} \left( Y_{2}^{-1} - Y_{2}^{1} \right), \\
	\psi_{d_{yz}} &= \frac{\mi}{\sqrt{2}} \left( Y_{2}^{-1} + Y_{2}^{1} \right), \\
	\psi_{d_{z^2}} &= Y_{2}^{0}, \\
	\psi_{d_{xy}} &= \frac{\mi}{\sqrt{2}} \left( Y_{2}^{-2} - Y_{2}^{2} \right), \\
	\psi_{d_{x^2 - y^2}} &= \frac{1}{\sqrt{2}} \left( Y_{2}^{-2} + Y_{2}^{2} \right) ,
\end{align}
\end{subequations}
we find that each wave function remains invariant under the time-reversal symmetry operation, as all of them are purely real.

For the spin component, the time-reversal symmetry operation is given by
\begin{equation}
	\hat{\Theta}^{-1} H \hat{\Theta} = (\mi \Sigma_y)^{-1} H \mi \Sigma_y \ \ \ , \ \ \ \mi \Sigma_y = (\mi \sigma_y) \oplus (\mi \sigma_y) \oplus \cdots,
\end{equation}
where $\mi \Sigma_y$ is the direct plus of $\mi \sigma_y$  across the spin subspace of each orbital. 

By combining both parts, we obtain the Hamiltonian after the time-reversal symmetry operation ${H}^\prime={\hat{\Theta}}^{-1}H\hat{\Theta}$. However, the new Hamiltonian is still not yet TRS because ${\hat{\Theta}}^{-1}H^\prime\ \hat{\Theta} \neq H$. To enforce TRS, we define a new Hamiltonian $\bar{H}$ as
\begin{equation}
	\bar{H} = \frac{H + \hat{\Theta}^{-1} H \hat{\Theta}}{2}.
\end{equation}
It is straightforward to verify that $\bar{H}$ is time-reversal symmetric by showing
\begin{equation}
	\hat{\Theta}^{-1} \bar{H} \hat{\Theta} = \hat{\Theta}^{-1} \left(\frac{H + \hat{\Theta}^{-1} H \hat{\Theta}}{2}\right) \hat{\Theta} = \frac{1}{2} \left(\hat{\Theta}^{-1} H \hat{\Theta} + \hat{\Theta}^{-1} \hat{\Theta}^{-1} H \hat{\Theta} \hat{\Theta} \right)
	= \frac{1}{2} \left(\hat{\Theta}^{-1} H \hat{\Theta} + H \right) = \bar{H}.
\end{equation}

In practice, we use this final Hamiltonian $\bar{H}$ in subsequent calculation. To claim that the effect on the structure of Hamiltonian is tiny during the operation, we define a quantity $\eta(H)$ as 
\begin{equation}
	\eta (H) = \mathrm{Tr} (H * \bar{H}) / \sqrt{\mathrm{Tr} (H * H) * \mathrm{Tr} (\bar{H} * \bar{H})},
\end{equation}
which is valued more than 0.998 in our calculation, so the operation is believed.

\subsection{Angular momentum projection}
\subsubsection{Orbital angular momentum (OAM) projection}

For OAM projection calculation, we consider the contributions from each orbital separately. Firstly, for the $s$-orbital, there is no contribution to the OAM projection,as its orbital quantum number is $l=0$ and magnetic quantum number $m=0$. 

For the $p$-orbitals, one assumes the calculated wave function $\left|\phi_p\right\rangle$ given by VASP has the form
\begin{equation}
	\left|\phi_p\right\rangle=\left(\alpha_{r}+\mathrm{i} \alpha_{i}\right)\left|p_{x}\right\rangle+\left(\beta_{r}+\mathrm{i} \beta_{i}\right)\left|p_{y}\right\rangle+\gamma\left|p_{z}\right\rangle, 
\end{equation}
where $\alpha_r$, $\alpha_i$ are the real and imaginary parts of coefficients of $\left|p_x\right\rangle$ term in $\left|\phi_p\right\rangle$, also, $\beta_r$, $\beta_i$ are the the real and imaginary parts for $\left|p_y\right\rangle$, and $\gamma$ is the purely real coefficient for $\left|p_z\right\rangle$ due to the symmetry from spherical harmonic function. Then the expected value of OAM yields 
\begin{equation}
	\left\langle\phi_p\right| \hat{L}_z\left|\phi_p\right\rangle=2\left(\alpha_r \beta_i-\beta_r \alpha_i\right).
\end{equation}

Similarly, for the $d$-orbitals, denoted the calculated wave function as $\left|\phi_d\right\rangle$, and assumed the coefficients taking the form
\begin{equation}
	\left|\phi_d\right\rangle=\left(\alpha_r+\mathrm{i} \alpha_i\right)\left|d_{xy}\right\rangle+\left(\beta_r+\mathrm{i} \beta_i\right)\left|d_{yz}\right\rangle+\left(\gamma_{r}+\mathrm{i} \gamma_{i}\right)\left|d_{xz}\right\rangle+\left(\delta_{r}+\mathrm{i} \delta_{i}\right)\left|d_{x^2-y^2}\right\rangle+\eta\left|d_{z^2}\right\rangle,
\end{equation}
we can similarly get the contribution to the OAM from this state is given by
\begin{equation}
	\left\langle\phi_d\right| \hat{L}_z\left|\phi_d\right\rangle=4\left(\alpha_i \delta_r-\alpha_r \delta_i\right)+2\left(\beta_r \gamma_r-\beta_r \gamma_i\right).
\end{equation}

Due to the orthogonality of spherical harmonic function, the contributions to OAM should be linear summed over all of orbitals. Therefore, we obtain the final OAM projection as
\begin{equation}
	\langle \hat{L}_z \rangle = \left\langle\phi_d\right| \hat{L}_z\left|\phi_d\right\rangle + \left\langle\phi_d\right| \hat{L}_z\left|\phi_d\right\rangle
\end{equation}

\subsubsection{Spin angular momentum (SAM) projection}

The SAM projection calculation is more straightforward than the OAM projection, as the projected magnetizations can be obtained directly from VASP. The SAM projection is given by
\begin{equation}
	\langle \hat{S}_{j=x,y,z} \rangle = \frac{1}{2} \sum_{n, k} \sum_{l, m} \sum_{\mu, v=\uparrow, l} \sigma_{\mu v}^j\left\langle\chi_{n k}^\mu | Y_l^m\right\rangle\left\langle Y_l^m | \chi_{n k}^v\right\rangle
\end{equation}
where $Y_l^m$ is the spherical harmonic with orbital and magnet quantum numbers $l$ and $m$, $\sigma_{\mu\nu}^j$ is the matrix element of the Pauli matrix for the spin basis $\mu,\nu=\uparrow,\downarrow$. Here, $j=x,y,z$ refers to the Pauli matrices for spin-1/2
\begin{equation}
 	\sigma^x=\left(\begin{array}{ll}
	0 & 1 \\
	1 & 0
	\end{array}\right) \ \ \ , \ \ \ \sigma^y=\left(\begin{array}{cc}
	0 & -\mi \\
	\mi & 0
	\end{array}\right) \ \ \ , \ \ \ \sigma^z=\left(\begin{array}{cc}
	1 & 0 \\
	0 & -1
	\end{array}\right)
\end{equation} 
and $\left|\chi_{nk}^{\mu,\nu}\right\rangle$ is a spinor corresponding to $n$-th energy band and momentum coordinate $\bm{k}$.

For our SAM projection, we evaluate the projected magnetizations for $j=z$ case and sum over each orbital.

\subsection{Green function method}
\subsubsection{Landauer-B{\"u}ttiker formula}

The Landauer-Büttiker (LB) method describes the relationship between the wave functions of electrons in a quantum junction and the conductance of that junction, essentially serving as a scattering-based approach~\cite{ryndyk2016theory,datta1997electronic}. In the LB method, the total system is divided into three parts: a left (equilibrium) electrode, a scattering region and a right (equilibrium) electrode, where scattering occurs among these regions. In the case of finite temperature and voltage across the electrodes, the Landauer formula~\cite{landauer1957spatial,landauer1970electrical} is given by
\begin{equation}
	I = I_{L \rightarrow R} - I_{R \rightarrow L} = \frac{e}{h} \int \md E \ T (E) [f_L (E) - f_R (E)]
	\label{Landauer_single-channel}
\end{equation}
where $E$ and $V=\varphi_L-\varphi_R$ denote energy of ingoing electrons and voltage between left and right electrodes, respectively. Here, $e$ and $h$ are the electron charge and Planck constant, $T\left(E,V\right)$ is the transmission function at given $E$ and $V$, and $f_L\left(E\right)$, $f_R\left(E\right)$ are Fermi-Dirac distribution functions
\begin{equation}
	f_{s = L, R} (E) = 1 / \{\exp[{(E - \mu_s - e \varphi_s) / T_s}] + 1\},
\end{equation}
where $\mu_s$ is the chemical potential of electrode $s$ (typically equal to the Fermi energy, $E_F$), and $T_s$ is the temperature of electrode $s$.

Büttiker extended this to a general Landauer formula for multi channels~\cite{PhysRevLett.57.1761,buttiker1988symmetry,PhysRevB.41.7906}
\begin{equation}
	I_s(V)=\frac{e}{h} \sum_p \int_{-\infty}^{\infty} d E\left[T_{p s}(E, V) f_s(E)-T_{s p}(E, V) f_p(E)\right],
	\label{L-B formula}
\end{equation}
where $s$ and $p$ denote channels indices. 

\subsubsection{Equilibrium Green function method}

We use matrix Green function method, which is commonly used to calculation the transport properties of mesoscopic systems~\cite{datta1997electronic,ryndyk2016theory}. The matrix Green function is defined as
\begin{equation}
	[(E + \mi \eta) \hat{\mathbb{I}} - \bm{H}] \bm{G}^{\mathrm{R}} = \hat{\mathbb{I}} \ \ \ \Rightarrow \ \ \ 
	\bm{G}^{\mathrm{R}} = [(E + \mi \eta) \hat{\mathbb{I}} - \bm{H}]^{-1},
\end{equation}
where $\eta$ is an infinitely small constant, approaching zero at the end of the calculation to avoid divergence of Green functions, and $\hat{\mathbb{I}}$ represents identity matrix. 

To consider the the system divided by left and right electrodes (denoted as L and R) and the scattering region (denoted as C), one can represent the Hamiltonian and Green function in matrix form as
\begin{equation}
	\bm{H} = 
	\begin{pmatrix}
	\bm{H}_L & \bm{V}_{LC} & 0 \\
	\bm{V}_{CL} & \bm{H}_C & \bm{V}_{CR} \\
	0 & \bm{V}_{RC} & \bm{H}_R
	\end{pmatrix} 
	\ \ \ , \ \ \ 
	\bm{G} = 
	\begin{pmatrix}
	\bm{G}_L & \bm{G}_{LC} & 0 \\
	\bm{G}_{CL} & \bm{G}_C & \bm{G}_{CR} \\
	0 & \bm{G}_{RC} & \bm{G}_R
	\end{pmatrix}	
\end{equation}
where assuming Hamiltonian is hermitian (represented in coupling part $\bm{V}$), given by $\bm{V}_{CL} = \bm{V}_{LC}^{\dagger}$ and $\bm{V}_{CR} = \bm{V}_{RC}^{\dagger}$. Noted that, we assume there is no coherence between the left and right electrodes, resulting in zero terms for the $LR$ indices in both Hamiltonian and Green function. 

It can be solved as~\cite{ryndyk2016theory,datta1997electronic}
\begin{align}
	\bm{G}_C &= [(E + \mi \eta) \hat{\mathbb{I}} - \bm{H}_C - \bm{\Sigma}]^{-1}, \\ 
	\bm{\Sigma} & 
	%=\bm{V}_{LC}^{\dagger} [(E + \mi \eta) \hat{\mathbb{I}} - \bm{H}_L]^{-1} \bm{V}_{LC} + \bm{V}_{RC}^{\dagger} [(E + \mi \eta) \hat{\mathbb{I}} - \bm{H}_L]^{-1} \bm{V}_{RC}
	= \bm{V}_{LC}^{\dagger} \bm{G}^\mathrm{R}_L \bm{V}_{LC} + \bm{V}_{RC}^{\dagger} \bm{G}^\mathrm{R}_R \bm{V}_{RC},
\end{align}
where $\bm{\Sigma}$ is the \textbf{self-energy} contribution from the electrodes, and evluating out
\begin{align}
	%\bm{G}^\mathrm{R}_{s = L, R} &= [(E + \mi \eta) \hat{\mathbb{I}} - \bm{H}_{s}]^{-1} \\ 
    \bm{\Sigma}^\mathrm{R}_L &= \bm{V}_{LC}^\dagger \bm{G}^{\textrm{R}}_L \bm{V}_{LC}, \\
    \bm{\Sigma}^\mathrm{R}_R &= \bm{V}_{RC}^\dagger \bm{G}^{\textrm{R}}_R \bm{V}_{RC},
\end{align}
with $\bm{G}^{\textrm{R}}_{L,R}$ being the retarded surface Green function of left and right electrodes, respectively. 

By subsitituing the wave functions into Schr{\"o}dinger equation, it can be derived out the transmission coefficient~\cite{paulsson2002non} 
\begin{equation}
	T(E) = \tr [\bm{\Gamma}_L (E) \bm{G}_C^{\text{A}} (E) \bm{\Gamma}_R (E) \bm{G}_C^{\text{R}} (E) ]
\end{equation}
where the level-width function $\bm{\Gamma}_{s=L,R}$ are defined as
\begin{equation}
	\bm{\Gamma}_{s=L,R} = \mi \left(\bm{\Sigma}_s^{\text{R}} - \bm{\Sigma}_s^{\text{A}}\right)
\end{equation}

Combining it with Landauer-B{\"u}tiker fomula, we an evaluate out the results of properties based on (equilibrium) Green function method, 
\begin{equation}
	I(V)=\frac{e}{h} \int_{-\infty}^{\infty} \mathrm{d} E \ \mathrm{Tr}\left[\bm{\Gamma}_L(E) \bm{G}_C^{\mathrm{A}}(E) \bm{\Gamma}_R(E) \bm{G}_C^{\mathrm{R}}(E)\right]\left[f_L(E)-f_R(E)\right].
	\label{EGF-equation}
\end{equation}

\subsubsection{Non-equilibrium Green function (NEGF) method}

The non-equilibrium Green function (NEGF) method is a more powerful and general approach for describing quantum transport properties at nanoscale, providing advantages over the equilibrium Green function method. 
It comes from Meir-Wingreen formula~\cite{PhysRevLett.68.2512,PhysRevB.50.5528,caroli1971direct}
\begin{equation}
	I_{s=L,R} (E) = \frac{\mi e}{\hslash} \int \md E \ \tr  [\bm{\Gamma}_s \bm{G}_C^{<} + f_s (E) \bm{\Gamma}_s (\bm{G}_C^{\text{R}} - \bm{G}_C^{\text{A}})]
\end{equation}
and get the $I$-$V$ relation~\cite{datta1997electronic,PhysRevLett.68.2512,PhysRevB.50.5528,caroli1971direct,keldysh1964diagram,keldysh1965diagram,PhysRev.150.516}
\begin{equation}
	I (E) = \frac{e}{\hslash} \int \md E \ [f_L (E - e \varphi) - f_R (E)] \mathrm{Tr} (\bm{\Gamma}_L \bm{G}_C^{\text{R}} \bm{\Gamma}_R \bm{G}_C^{\text{A}}),
	\label{NEGF-equation}
\end{equation}
which is equivalent to Eq.\eqref{EGF-equation}. We will use this formula in subsequent $I$-$V$ calculations.

\subsubsection{Calculation details}

Our calculations are based on Eq.\eqref{EGF-equation}, with an applied voltage $\varphi$ between left and right leads. The $I$-$V$ formula becomes~\cite{PhysRevB.73.085414}
\begin{equation}
	T(E) = \mathrm{Tr} [\bm{\Gamma}_L (E - e \varphi / 2) \bm{G}_C^{\text{R}} (E) \bm{\Gamma}_R (E + e \varphi / 2) \bm{G}_C^{\text{A}} (E)] 
	\ \  \text{, where} \ \  
	\bm{G}_C^{\text{R}} (E) = [(E - E_F + \mi \eta) \hat{\mathbb{I}} - \bm{H}_C - \bm{\Sigma}]^{-1}.
\end{equation}
We apply tight-binding Hamiltonian models for a CNAC with Wannier-basis Hamiltonian. To make the effects of boundaries and interface between leads and systems less, we consider itself as left and right leads. Here, VASP and Wannier90 can calculate out the Hamiltonian of CNAC itself, and the coupling term between two neighboring CNAC. It needs to noted that the tight-binding Hamiltonian is incorporated time-reversal symmetry as discussed above. 

\subsection{Time-dependent calculations}

To investigate light-induced magnetization dynamics, we utilize time-dependent spin density functional theory (TDSDFT) within the Kohn-Sham (KS) formalism~\cite{theilhaber1992ab, yabana1996time, xu2024real, siegrist2019light}. The system’s ground state is first obtained using spin-polarized DFT and subsequently evolved in real-time by applying the following motion equations:
\begin{equation}
\mi \partial_t\left|\psi_{n, \bm{k}}^{\text{KS}}(t)\right\rangle=\left\{\frac{1}{2}\left[-\mi \nabla+\frac{A(t)}{c}\right]^2 \sigma_0+V_{\text{KS}}(t)\right\}\left|\psi_{n, \bm{k}}^{\text{KS}}(t)\right\rangle ,
\end{equation}
where $\left|\psi_{n, \bm{k}^{\text{KS}}}(t)\right\rangle$ represents the KS-Bloch state at $k$-point $\bm{k}$ and band index $n$, defined as a Pauli spinor:
\begin{equation}
\left|\psi_{n, \bm{k}}^{\text{KS}}(t)\right\rangle=\left[\begin{array}{c}
\left|\psi_{n, \bm{k}, \uparrow}^{\text{KS}}(t)\right\rangle \\ [0.5em]
\left|\psi_{n, \bm{k}, \downarrow}^{\text{KS}}(t)\right\rangle
\end{array}\right].
\end{equation}
Here, $\left|\psi_{n, \bm{k}, \alpha=\uparrow, \downarrow}^{\text{KS}}(t)\right\rangle$ denotes the spin-up/spin-down components, with $\sigma_0$ as the $2 \times 2$ identity matrix and $\bm{A}(t)$ as the vector potential of the incident laser pulse in the dipole approximation, where $-\partial_t \bm{A}(t) = c\bm{E}(t)$, and $c$ is the speed of light. The time-dependent KS potential $V_{\text{KS}}(t)$ is defined as:
\begin{equation}
	V_{\text{KS}}(t) = \int \md^3 r^{\prime} \ \frac{n\left(r^{\prime}, t\right)}{\left|r-r^{\prime}\right|} \sigma_0+V_{X C}[\rho(r, t)]+V_{ion}.
\end{equation}
The first term is the classical Hartree term, representing the mean-field electrostatic interaction, where \\ $n(\bm{r}, t)=\sum_{n, \bm{k}, \alpha} W_{\bm{k}}\left|\left\langle \bm{r} \mid \varphi_{n, \bm{r}, \alpha}^{\text{KS}}(t)\right\rangle\right|^2$ is the time-dependent electron density. $V_{XC}$ denotes the exchange-correlation potential, dependent on the spin density matrix $\rho(\bm{r}, t)$ within the local spin density approximation:
\begin{equation}
	\rho(\bm{r}, t)=\frac{1}{2} n(\bm{r}, t) \sigma_0+\frac{1}{2} \bm{m}(\bm{r}, t) \cdot \bm{\sigma} ,
\end{equation}
where $\bm{m}(\bm{r}, t)$ is the time-dependent magnetization vector: 
\begin{equation}
	\bm{m}(\bm{r}, t)=\sum_{n, \bm{k}} W_{\bm{k}}\left\langle\psi_{n, \bm{k}}^{\text{KS}}(t) \mid \bm{r}\right\rangle \bm{\sigma}\left\langle\bm{r} \mid \varphi_{n, \bm{k}}^{\text{KS}}(t)\right\rangle .
\end{equation}
The term $V_{ion}$ denotes electron interactions with lattice ions and core electrons, with the frozen core approximation employed to minimize computational costs. The bare Coulomb potential is replaced by a fully relativistic nonlocal norm-conserving pseudopotential~\cite{hartwigsen1998relativistic}, which includes spin-orbit coupling (SOC) terms proportional to $\bm{L} \cdot \bm{S}$, where $\bm{L} = (L_x, L_y, L_z)$ is the angular momentum operator vector and $\bm{S} = \frac{1}{2} \bm{\sigma} = \frac{1}{2} (\sigma_x, \sigma_y, \sigma_z)$ is the spin operator vector. with $\sigma_i$ the $i$-th Pauli matrix. It is noteworthy that $V_{\text{KS}}$ is nondiagonal in spin space due to the spin-orbit coupling term. 

The laser-electron interaction is expressed in the velocity gauge, with the vector potential $\bm{A}(t)$ defined as:
\begin{equation}
	\bm{A}(t)=f(t) \frac{c E_0}{\omega} \sin (\omega t) \hat{\bm{e}} ,
\end{equation}
where $E_0$ is the field amplitude, $\omega$ is the carrier frequency, and $\hat{\bm{e}}$ is a unit vector typically indicating elliptical polarization. In this setup, nuclear motion is ignored under the frozen nuclei approximation, deemed valid for heavy atoms at attosecond to femtosecond scales. Furthermore, due to the dipole approximation, magnetic components of the laser pulse are not included, as their impact is considered minimal in these driving conditions.

The KS equations of motion are resolved on a real-space grid using the Octopus code~\cite{castro2006octopus,andrade2015real,tancogne2020octopus}, enabling the calculation of time-dependent observables like total electronic current $\bm{J}(t) = \frac{1}{\Omega} \int \md r^3 \ \bm{j} (\bm{r}, t)$, where $\Omega$ is unit cell volume, and $\bm{j} (\bm{r}, t)$ is the microscopic time-dependent current density:
\begin{equation}
	\bm{j}(\bm{r}, t)=\sum_{n, \bm{k}, \alpha}\left[\varphi_{n, \bm{k}, \alpha}^{\text{KS \ *}}(\bm{r}, t)\left(\frac{1}{2}\left(-i \nabla+\frac{A(t)}{c}\right)+\left[V_{i o n}, \bm{r}\right]\right) \varphi_{n, \bm{k}, \alpha}^{\text{KS}}(\bm{r}, t)+c.c.\right]+\bm{j}_{\bm{m}}(\bm{r}, t) ,
\end{equation}
where $\bm{j}_{\bm{m}}(\bm{r}, t)$ is the magnetization current density, which after spatial integration vanishes and does not contribute to $\bm{J} (t)$. The spin expectation values are calculated as $\bm{S}(t) = \langle\psi_{n,\bm{k}}^{\text{KS}}(t)|\bm{S}|\psi_{n,\bm{k}}^{\text{KS}}(t)\rangle$, and are used to track the spin dynamics in the system. The resulting induced magnetization is given as $\langle S_x(t) \rangle$ in units of Bohr magneton (for the 1D structure, the periodicity is on the $x$-axis), and is summed over all atoms and occupied electronic states for a given unit cell. We refer to a nonmagnetic material as a material with strictly $\langle \bm{S}(t=0)\rangle=0$, and where the spin expectation values vanish even when projecting on individual atomic sites in the unit cell. 

% \subsection{Supplemental figures}

\begin{figure}[!htbp]
\centering
\includegraphics[width=0.80\textwidth]{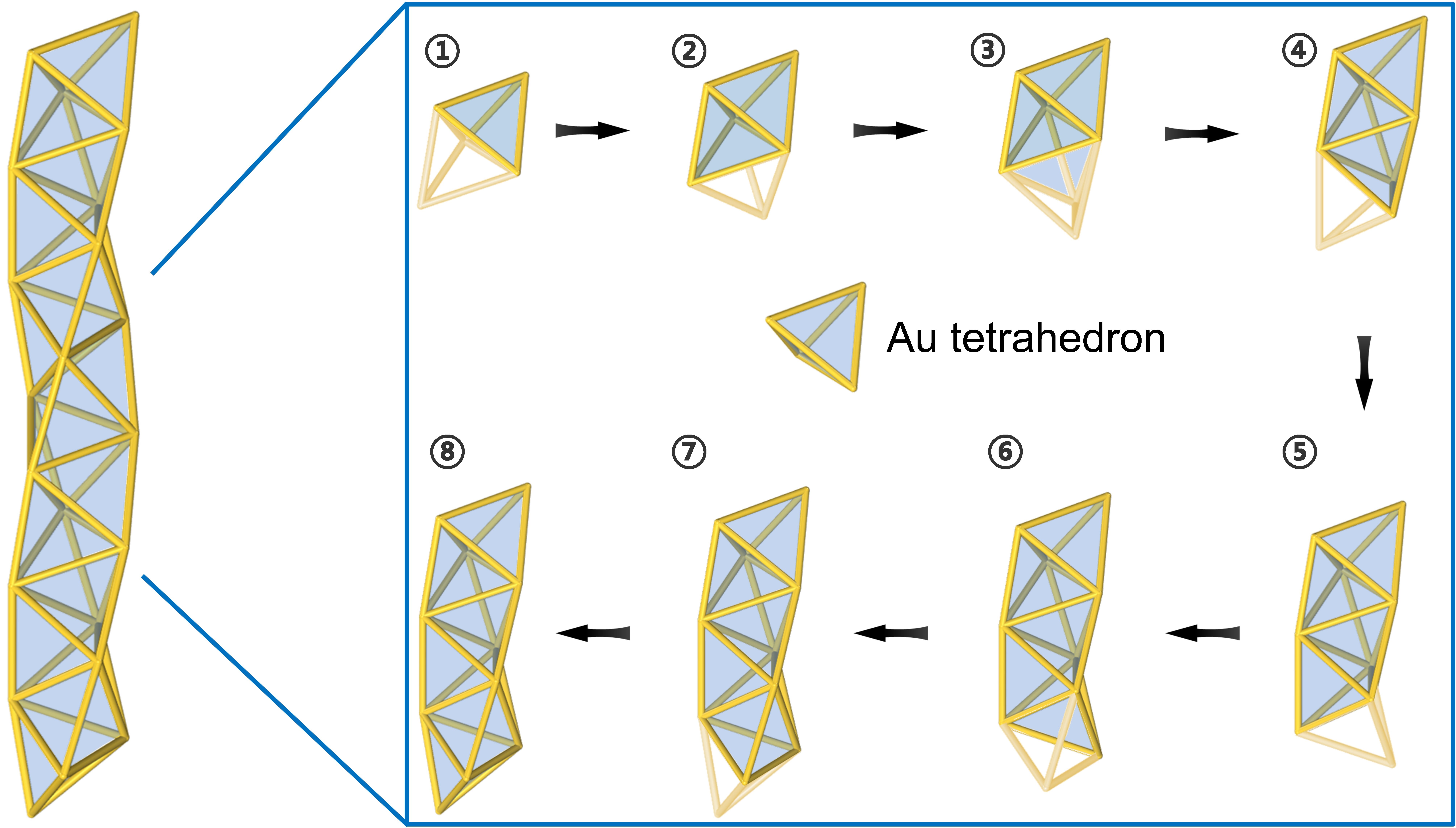}
\caption{\justifying
Schematic representation of the atomic model construction of L-CANC.
}
\label{fig:S1}
\end{figure}

\begin{figure}[!htbp]
\centering
\includegraphics[width=0.80\textwidth]{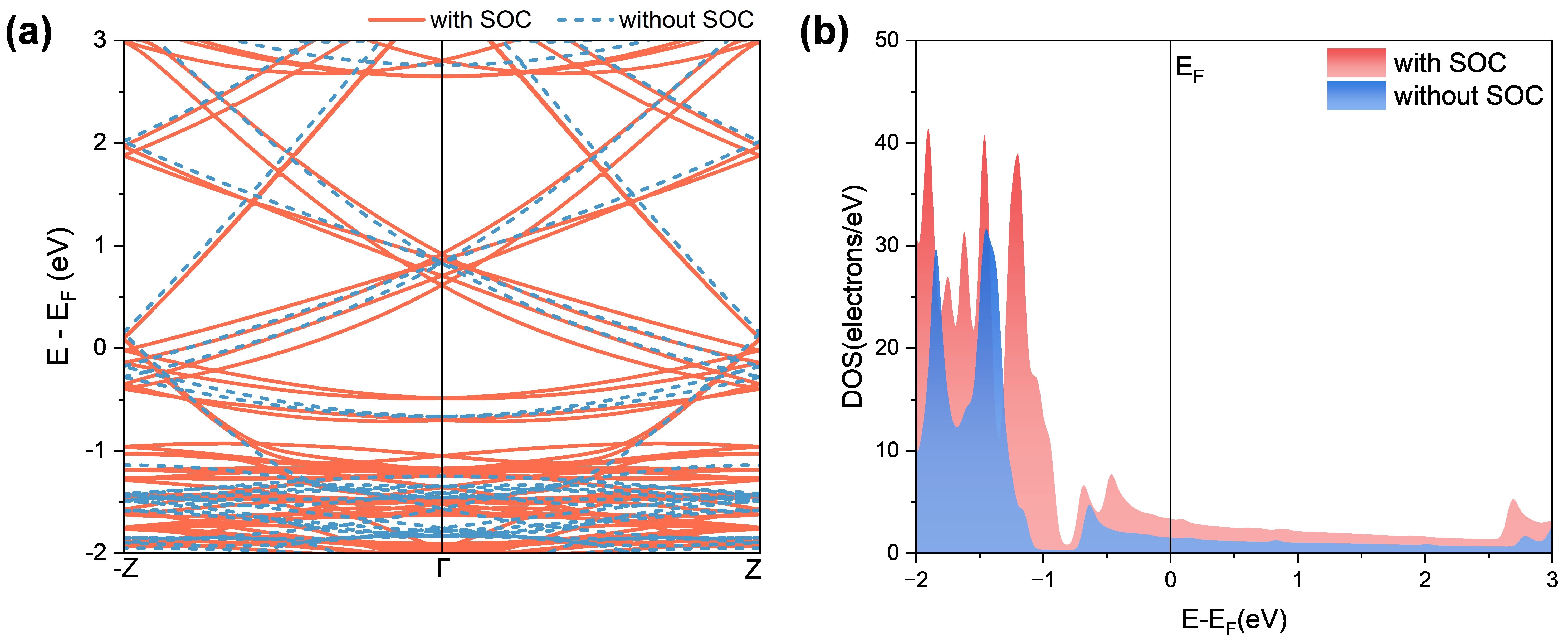}
\caption{\justifying
Ground-state electronic properties with and without SOC of L-CNAC. (a) \textit{ab-initio} band structure. (b) Density of states (DOS).
}
\label{fig:S2}
\end{figure}

\begin{figure}[!htbp]
\centering
\includegraphics[width=0.98\textwidth]{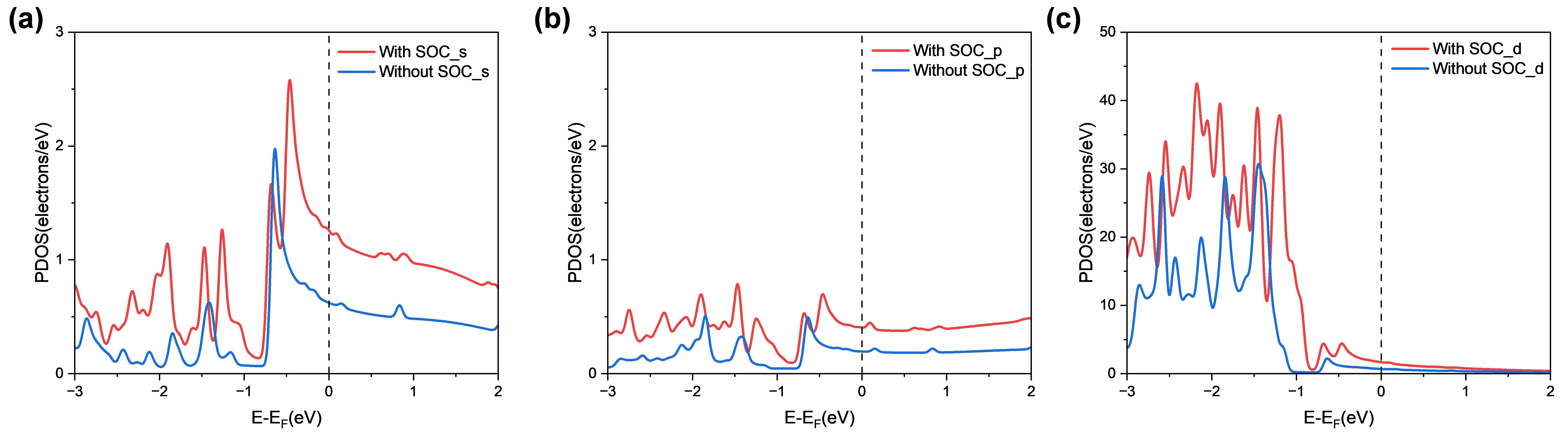}
\caption{\justifying
Projected density of states (PDOS) with and without SOC of L-CANC. (a) $s$-orbitals. (b) $p$-orbital. (c) $d$-orbitals. 
}
\label{fig:S3}
\end{figure}

\begin{figure}[!htbp]
\centering
\includegraphics[width=0.70\textwidth]{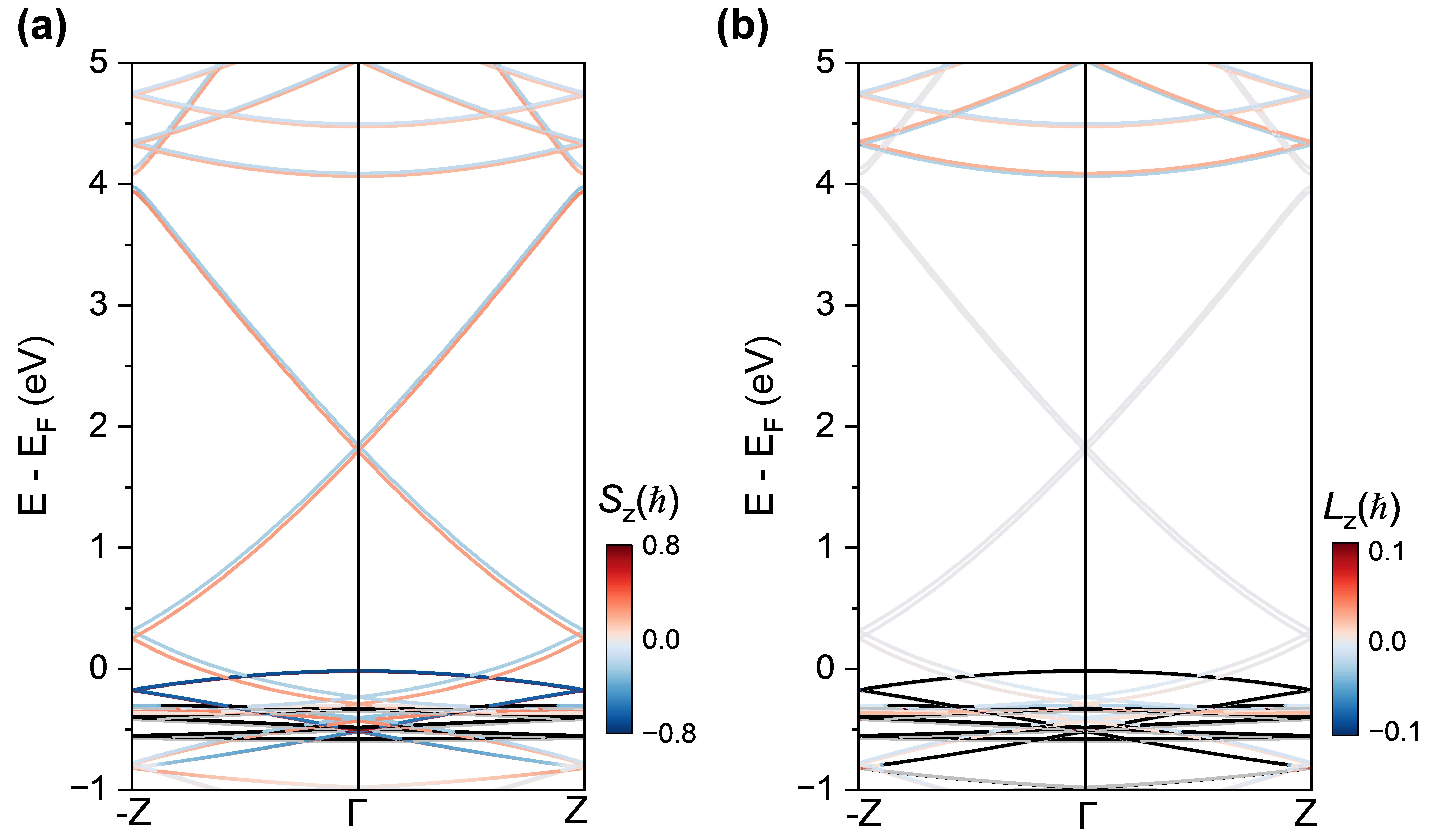}
\caption{\justifying
Spin projection and orbital texture of ANAC. (a) spin projection. (b) orbital texture.
}
\label{fig:S4}
\end{figure}

\begin{figure}[!htbp]
\centering
\includegraphics[width=0.98\textwidth]{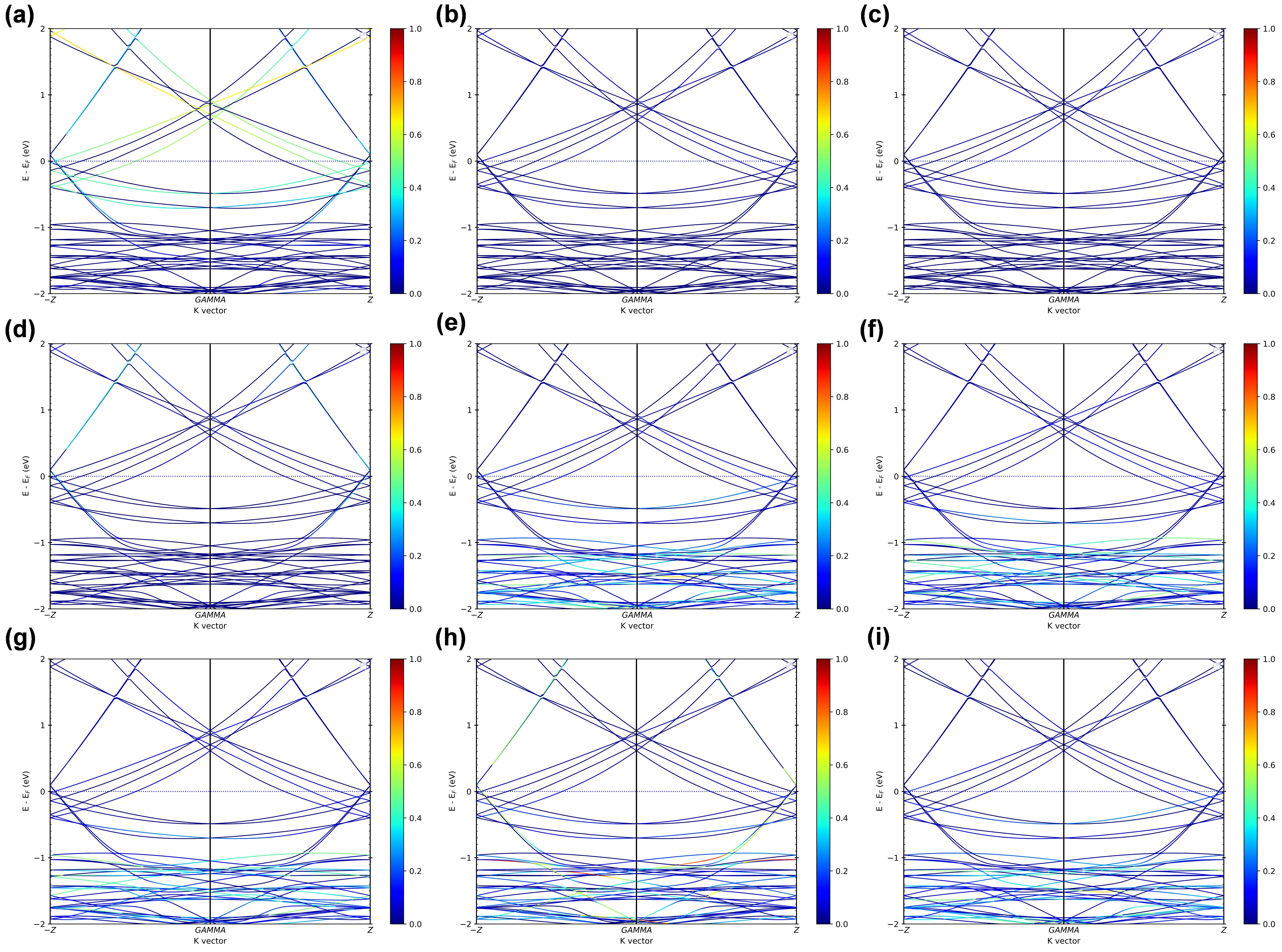}
\caption{\justifying
Fat bands of L-CNAC. (a) $s$-orbital. (b) $p_x$-orbital. (c) $p_y$-orbital. (d) $p_z$-orbital. (e) $d_{xy}$-orbital. (f) $d_{yz}$-orbital. (g) $d_{xz}$-orbital. (h) $d_{zz}$-orbital. (i) $d_{x^2-y^2}$-orbital.
}
\label{fig:S5}
\end{figure}

\begin{figure}[!htbp]
\centering
\includegraphics[width=0.98\textwidth]{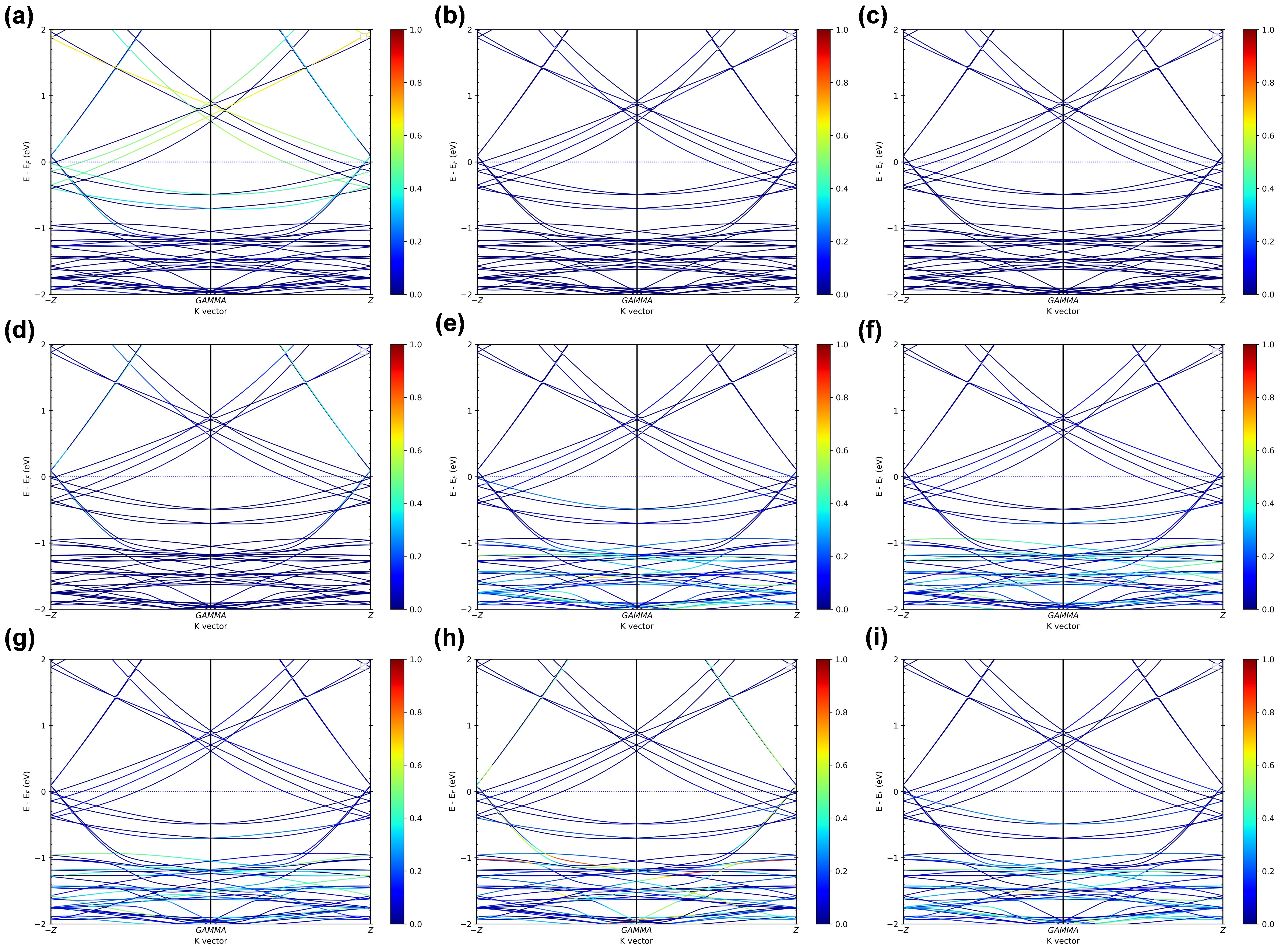}
\caption{\justifying
Fat bands of R-CNAC. (a) $s$-orbital. (b) $p_x$-orbital. (c) $p_y$-orbital. (d) $p_z$-orbital. (e) $d_{xy}$-orbital. (f) $d_{yz}$-orbital. (g) $d_{xz}$-orbital. (h) $d_{zz}$-orbital. (i) $d_{x^2-y^2}$-orbital.
}
\label{fig:S6}
\end{figure}

\begin{figure}[!htbp]
\centering
\includegraphics[width=0.98\textwidth]{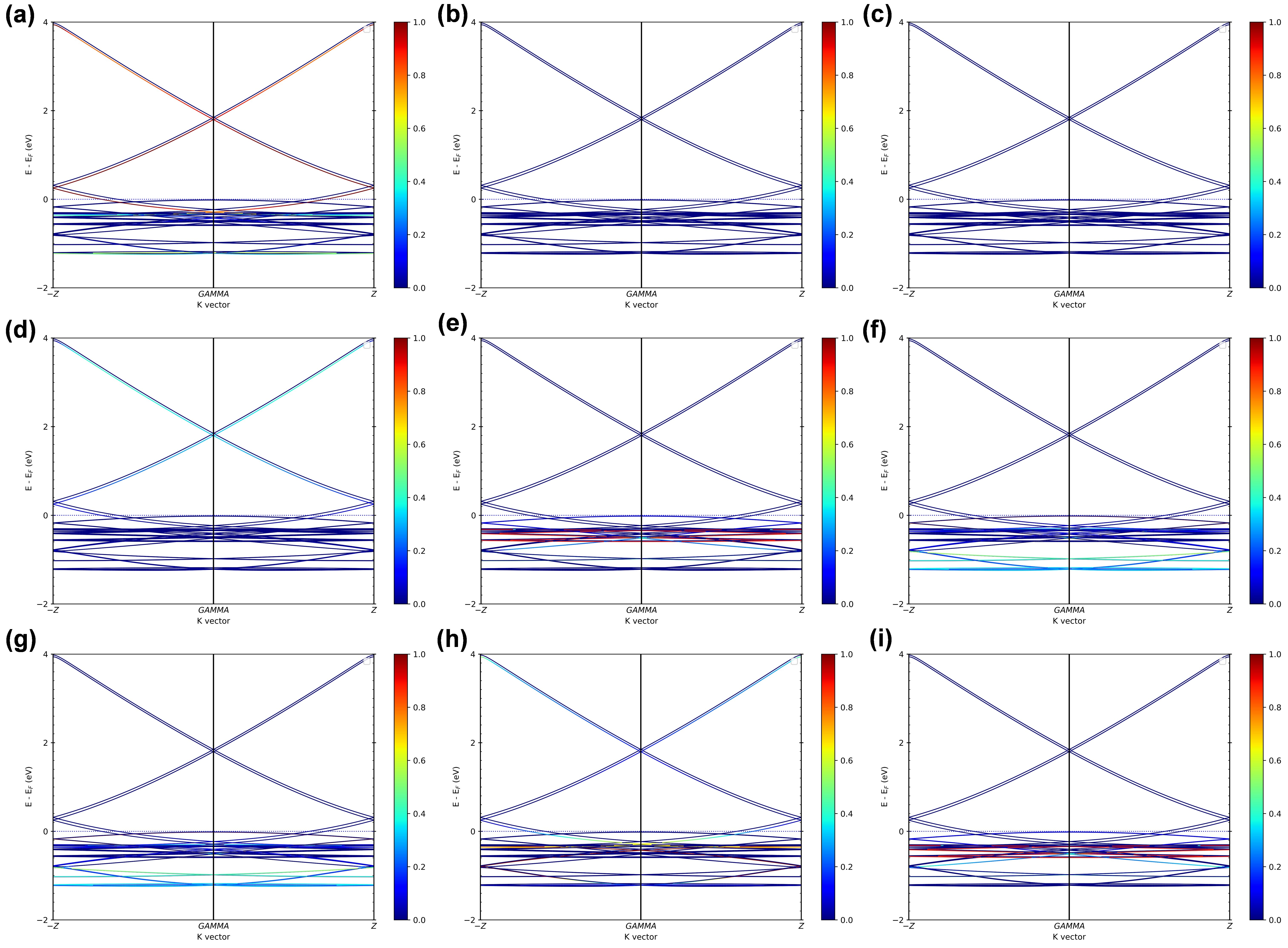}
\caption{\justifying
Fat bands of ACNAC. (a) $s$-orbital. (b) $p_x$-orbital. (c) $p_y$-orbital. (d) $p_z$-orbital. (e) $d_{xy}$-orbital. (f) $d_{yz}$-orbital. (g) $d_{xz}$-orbital. (h) $d_{zz}$-orbital. (i) $d_{x^2-y^2}$-orbital.
}
\label{fig:S7}
\end{figure}

\begin{figure}[!htbp]
\centering
\includegraphics[width=0.70\textwidth]{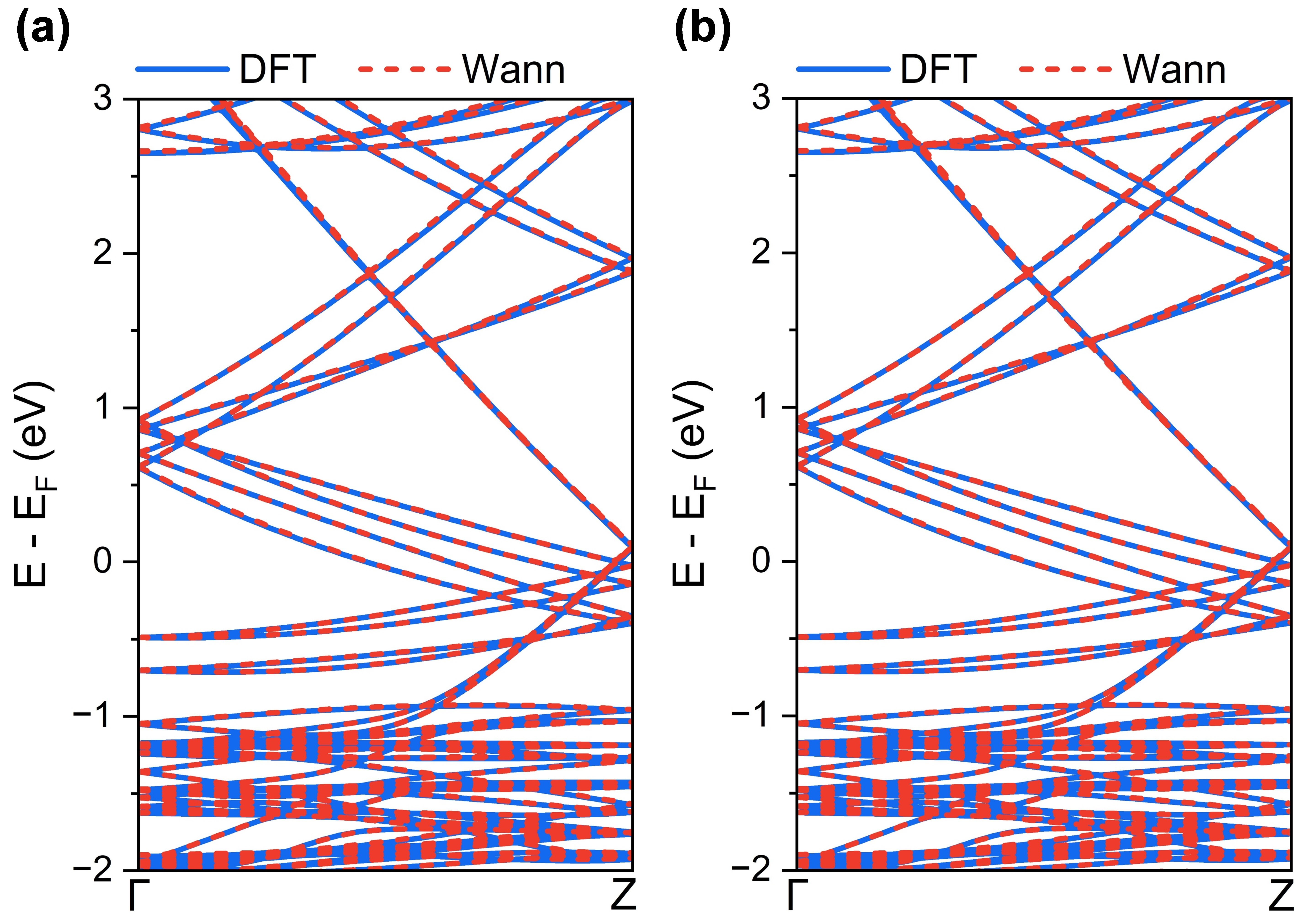}
\caption{\justifying
Wannier function and DFT band structures with SOC. (a) L-CNAC. (b) R-CNAC. 
}
\label{fig:S8}
\end{figure}

\begin{figure}[!htbp]
\centering
\includegraphics[width=0.90\textwidth]{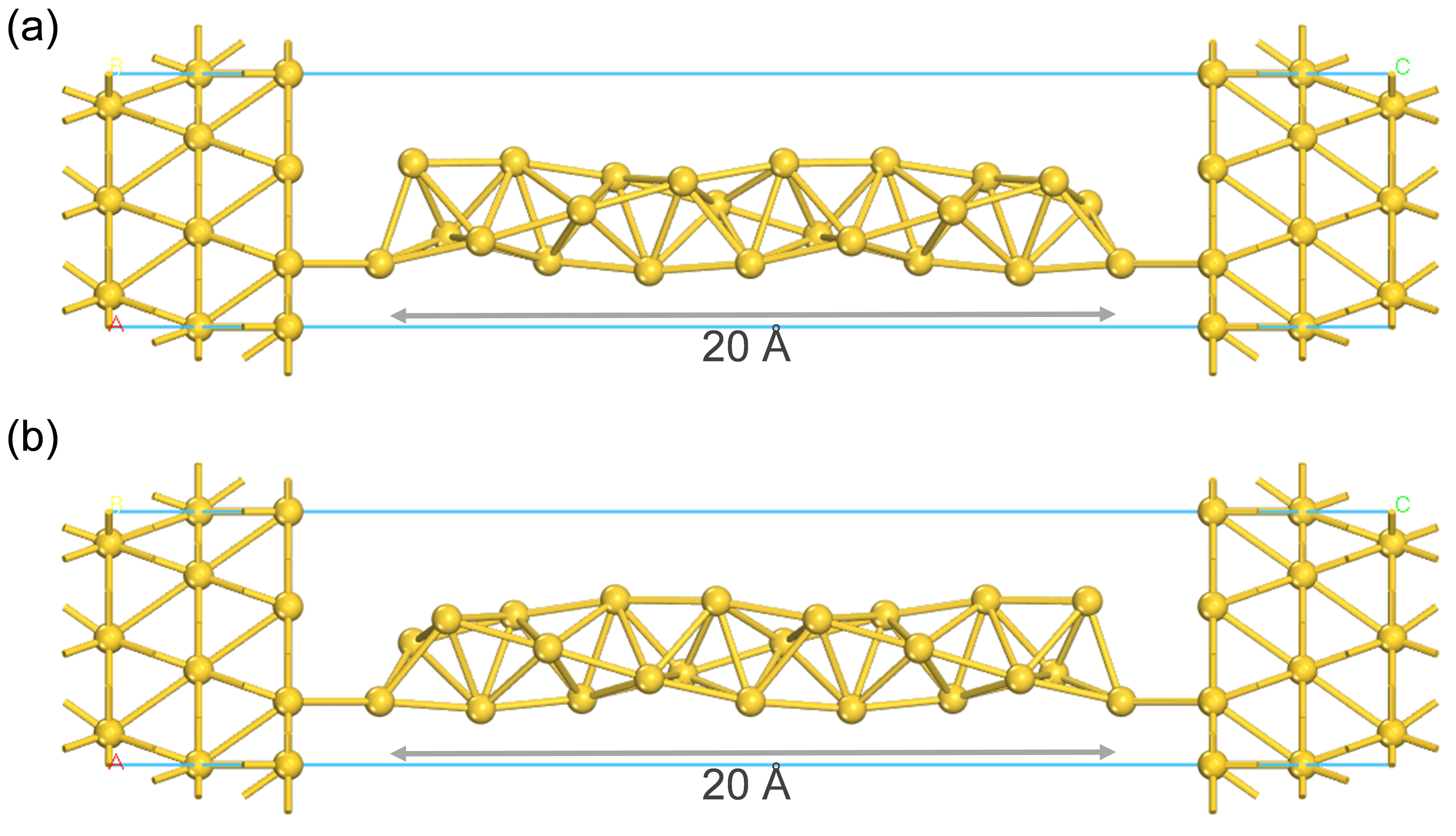}
\caption{\justifying
Device models. (a) L-CNAC. (b) R-CNAC.
}
\label{fig:S9}
\end{figure}

\begin{figure}[!htbp]
\centering
\includegraphics[width=0.80\textwidth]{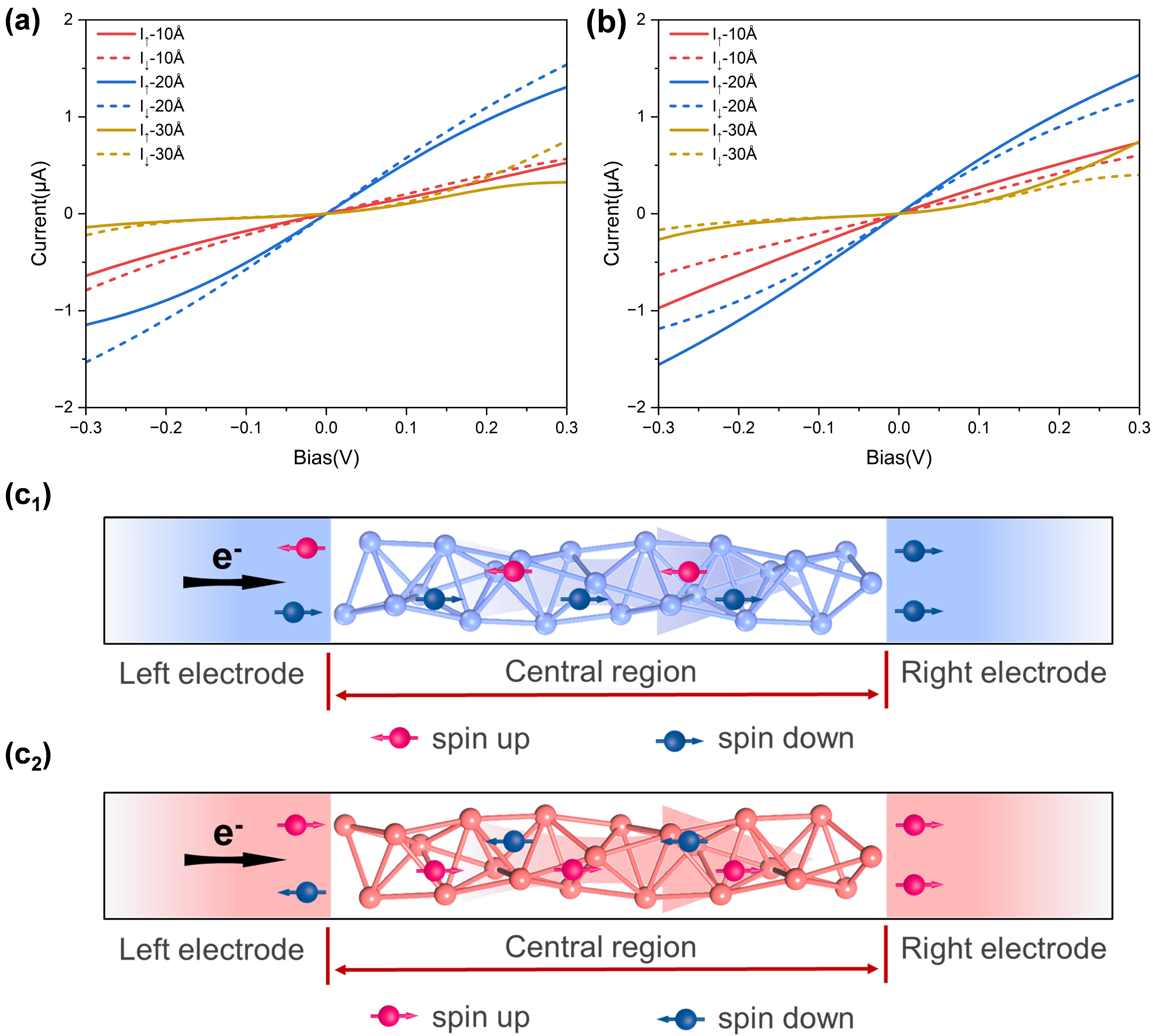}
\caption{\justifying
(a, b) The simulated current as a function of applied voltage for L-CNAC (a) and R-CNAC (b) device models. (c) Electrode with SOC, and the length of the central region approaches 10Å, 20Å, 30Å. Schematic illustrations of spin selective transport in a device model for L-CNAC (c$_1$) and RCNAC (c$_2$), where CNACs serve as the central region. 
}
\label{fig:S10}
\end{figure}

\begin{figure}[!htbp]
\centering
\includegraphics[width=0.45\textwidth]{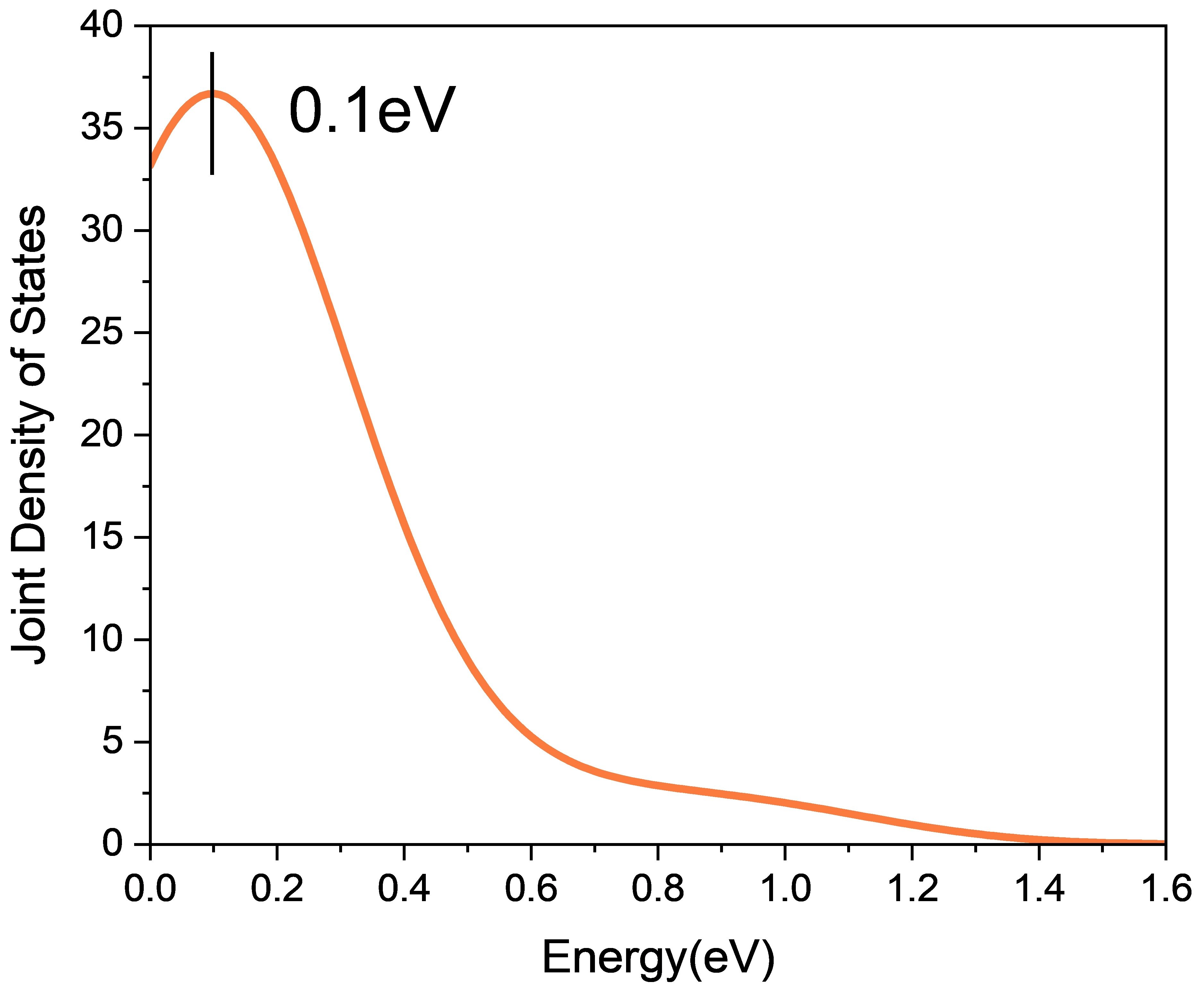}
\caption{\justifying
Joint density of states (JDOS) of L-CNAC. 
}
\label{fig:S11}
\end{figure}

\begin{figure}[!htbp]
\centering
\begin{minipage}{\textwidth} 
\centering
\includegraphics[width=0.98\textwidth]{FIGS12.jpg}
\caption{\justifying
Time evolution of chiral photomagnetic responsiveness with SOC. (a, b) Time-dependent occupations of spin-up and spin-down channels driven by 532nm linearly polarized light for L-CNAC (a) and R-CNAC (b). (c, d) Time-dependent magnetic moment components driven by 532nm linearly polarized light for L-CNAC (c) and R-CNAC (d). The $x$, $y$, and $z$ directions correspond to periodic boundary conditions for CNACs ($z$-axis).
}
\label{fig:S12}

\medskip 
% \justifying
\normalfont\normalsize\linespread{1}\selectfont\justifying
\indent The metallic character of CNACs, evidenced by high DOS near $E_F$, enables frequency-selective photomagnetic responses. JDOS analysis (Fig.~\ref{fig:S11}) and orbital-resolved excitation mapping (Figs.~\ref{fig:S5} and ~\ref{fig:S6}) confirm that 12400 nm excitation aligns with SOC-induced inter-band thresholds ($d$-orbital peaks at $E_F\pm 0.1$ eV), avoiding plasmonic absorption. Orbital-resolved transitions reveal $d_{xz}/p_z$ hybridization at the $Z$-point, generating $L_z\neq 0$ states where linearly polarized light preferentially excites $+L_z$ transitions. Crucially, 532 nm excitation (Fig.~\ref{fig:S12}) shows negligible response: spin polarization fluctuations ($<10^{-3}$$\mu$B) and random magnetic oscillations ($\pm0.01$$\mu$B) without chirality dependence, contrasting sharply with 160 fs stabilized moments under 12400 nm excitation Fig.~3(b, d). Geometric chirality converts orbital polarization into measurable spin currents and photomagnetic moments via SOC, overturning noble-metal photomagnetic limitations.
\end{minipage}
\end{figure}

\begin{figure}[!htbp]
\centering
\includegraphics[width=0.98\textwidth]{FIGS13.jpg}
\caption{\justifying
Time evolution of chiral photomagnetic responsiveness with SOC. (a, b) Time-dependent occupations of spin-up and spin-down channels driven by linearly polarized light at 100 fs for L-CNAC (a) and R-CNAC (b). (c, d) Time-dependent magnetic moment components driven by linearly polarized light at 100 fs for L-CNAC (c) and R-CNAC (d). The $x$, $y$, and $z$ directions correspond to periodic boundary conditions for CNACs ($z$-axis).
}
\label{fig:S13}
\end{figure}

\begin{figure}[!htbp]
\centering
\begin{minipage}{\textwidth} 
\includegraphics[width=0.98\textwidth]{FIGS14.jpg}
\caption{\justifying
Time evolution of chiral photomagnetic responsiveness with SOC. (a, b) Time-dependent occupations of spin-up and spin-down channels driven by linearly polarized light at 200 fs for L-CNAC (a) and R-CNAC (b). (c, d) Time-dependent magnetic moment components driven by linearly polarized light at 200 fs for  L-CNAC (c) and R-CNAC (d). The $x$, $y$, and $z$ directions correspond to periodic boundary conditions for CNACs ($z$-axis).
}
\label{fig:S14}

\medskip 
% \justifying
\normalfont\normalsize\linespread{1}\selectfont\justifying
\indent 
Our rt-TDDFT methodology employs the semiclassical light-matter interaction framework, treating the electromagnetic field classically while quantizing electronic degrees of freedom—$\alpha$ rigorously validated approach for conventional light intensities where quantum field fluctuations are negligible. This treatment remains distinct from quantum electrodynamical methods required for nonclassical light ($e.g.$, squeezed/entangled photons), which is rather challenging and time-consuming. The robustness of our semiclassical implementation is unequivocally demonstrated through systematic pulse-width comparisons, revealing virtually identical spin-flip dynamics [Figs. 3(g, h), ~\ref{fig:S13}(a, b), ~\ref{fig:S14}(a, b)] and magnetic moment evolution [Figs.~3(d, e), ~\ref{fig:S13}(c, d), ~\ref{fig:S14}(c, d)] across 100-300 fs excitations. These data confirm photomagnetic responses stabilize within 160 fs regardless of pulse duration, with SOC-driven chiral selectivity governed by $L_z$-momentum locking dominating sub-picosecond processes without pulse-width sensitivity. Critically, all magnetic components remain confined along the helical axis (z-direction per periodic boundary conditions), underscoring both the spatial specificity of these effects and the adequacy of semiclassical treatment for our system.
\end{minipage}
\end{figure} 

\begin{figure}[!htbp]
\centering
\begin{minipage}{\textwidth} 
\centering
\includegraphics[width=0.98\textwidth]{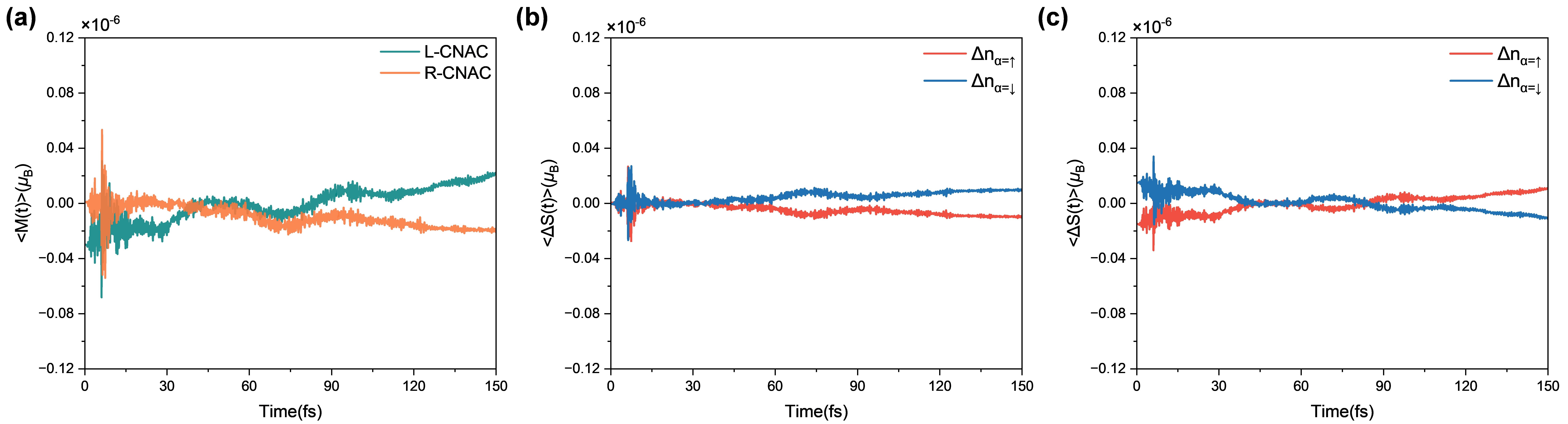}
\caption{\justifying
Time evolution of chiral photomagnetic responsiveness without SOC of CNAC. (a) Total magnetic moment variation in L-CNAC and R-CNAC under the same laser illumination, but without SOC. (b, c) Time-dependent occupations of spin-up and spin-down channels driven by linearly polarized light for L-CNAC (b) and R-CNAC (c), respectively. Note that the order of magnitude of the vertical axis is $10^{-6}$.
}
\label{fig:S15}

\medskip 
% \justifying
\normalfont\normalsize\linespread{1}\selectfont\justifying
\indent 
In Fig.~\ref{fig:S15}, the photomagnetic responsiveness of L-CNAC and R-CNAC under without SOC conditions is presented. Compared to the results in Fig. 3 (the case with SOC), the photomagnetic response of the entire system is significantly diminished by 6 orders of magnitude. The absence of coupling between spin states prevents the linearly polarized light from inducing significant magnetic changes. This highlights the significance of SOC in strong photomagnetic responses, especially in systems with chiral structures. The time-dependent spin channel occupation curves of L-CNAC and R-CNAC is also analyzed [Figs.~\ref{fig:S15}(b) and (c)], the amplitudes of $\Delta n_{\alpha=\uparrow}$ and $\Delta n_{\alpha=\downarrow}$ are extremely small, believed in the order of numerical calculation fluctuation. Under the lack of SOC, the responses of L-CNAC and R-CNAC are almost identical, showing that chiral structures alone are insufficient to induce photomagnetic responses. This means the coupling between the light field and magnetic response is significantly amplified if SOC is introduced, emerging in pronounced chiral-related photomagnetic effects.
\end{minipage}
\end{figure}

\begin{figure}[!htbp]
\centering
\begin{minipage}{\textwidth} 
\includegraphics[width=0.98\textwidth]{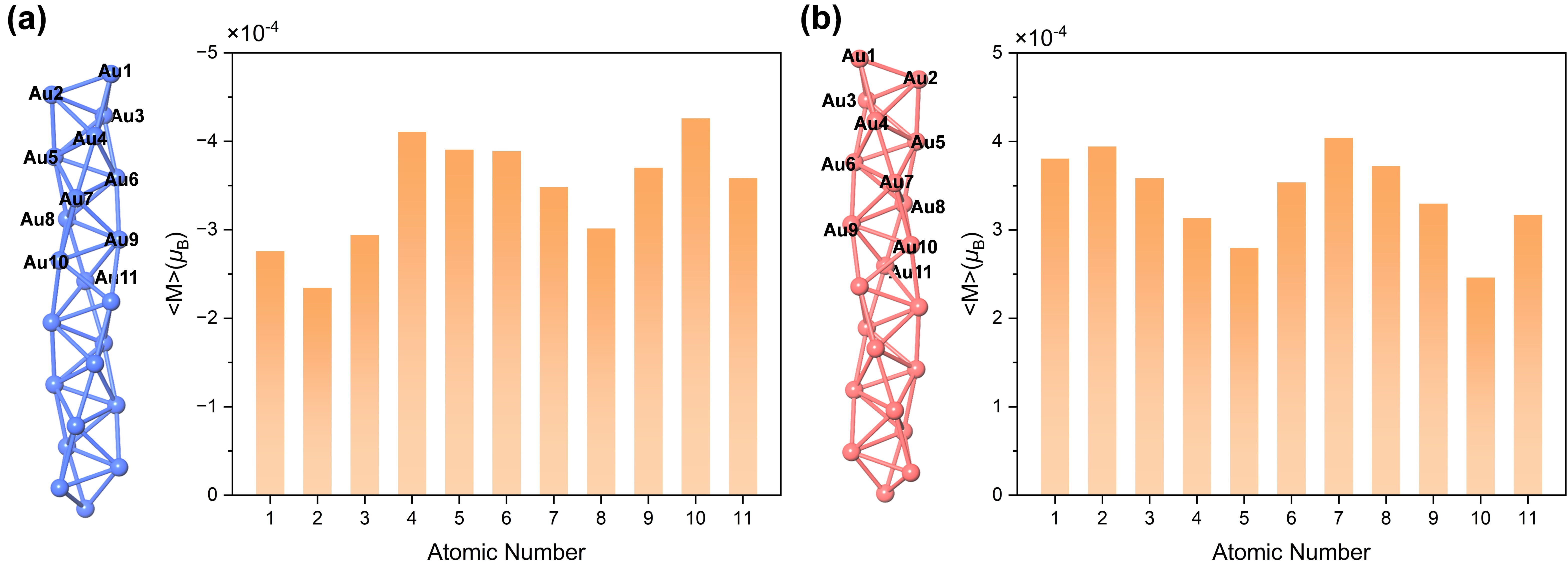}
\caption{\justifying
Time-dependent magnetic moment per atom (1 period) in L-CNAC (a) and R-CNAC (b) under linearly polarized light excitation at 160 fs.
}
\label{fig:S16}

\medskip 
% \justifying
\normalfont\normalsize\linespread{1}\selectfont\justifying
\indent 
Our method definitively resolves site-dependent spin polarization, as validated by Fig.~\ref{fig:S16}. The data reveals striking chiral asymmetry: in L-CNAC [Fig.~\ref{fig:S16}(a)], Au10 shows maximum peak magnetization ($\sim -4.5 \times 10^4$ $\mu$B), while the average magnetic moment of Au1-Au11 is around $\sim -2.5 \times 10^4$ $\mu$B. Conversely, R-CNAC [Fig.~\ref{fig:S16}(b)] exhibits a complete inversion asymmetry with L-CNAC. This non-uniform distribution directly originates from chiral-potential-modulated spin-orbit interactions $t$ specific atomic sites, demonstrating atomic-resolution capability for probing spin polarization landscapes.
\end{minipage}
\end{figure}

\end{document}